\documentclass[prb,showpacs,byrevtex,amsmath,amssymb,nofootinbib]{revtex4}
\usepackage{graphicx}
\usepackage{dcolumn}
\usepackage{bm}

\begin{document}
\preprint{PREPRINT}

\title[Short Title]{Like-Charge Colloid-Polyelectrolyte Complexation}

\author{Ren\'{e} Messina}
\email{messina@mpip-mainz.mpg.de}
\author{Christian Holm}
\email{holm@mpip-mainz.mpg.de}
\author{Kurt Kremer}
\email{k.kremer@mpip-mainz.mpg.de}
\affiliation{Max-Planck-Institut f\"{u}r Polymerforschung,
Ackermannweg 10, 55128 Mainz, Germany
}%

\date{\today}

\begin{abstract}
  We investigate the complexation of a highly charged sphere with a long
  flexible polyelectrolyte, \textit{both negatively charged} in salt free
  environment.  Electroneutrality is insured by the presence of divalent
  counterions.  Using molecular dynamics (MD) within the framework of the
  primitive model, we consider different Coulomb coupling regimes. At strong
  Coulomb coupling we find that the adsorbed chain is always confined to the
  colloidal surface but forms different conformations that depend on the
  linear charge density of the chain.  A mechanism involving the
  polyelectrolyte \textit{overcharging} is proposed to explain these
  structures.  At intermediate Coulomb coupling, the chain conformation starts
  to become three-dimensional, and we observe multilayering of the highly
  charged chain while for lower charge density the chain wraps around the
  colloid.  At weak Coulomb coupling, corresponding to an aqueous solvent, we
  still find like-charge complexation. In this latter case the chain
  conformation exhibits loops.
\end{abstract}

\maketitle

%
%
%

\section{Introduction\label{sec.Intro}}

The adsorption of polyelectrolytes onto an \textit{oppositely} charged
spherical particle has been experimentally extensively studied recently
\cite{Holde_chromatin_1989,McQuigg_JPhysChem_1992,Grunstein_Nature_1997,Caruso_Science_1998}.
Various authors have investigated this phenomenon theoretically
\cite{Muthukumar_JCP_1994,Marky_JMolBiol_1995,Sens_PRL_1999,Mateescu_EPL_1999,Park_EPL_1999,
Netz_Macromol_1999,Schiessel_EPL_2000,Kunze_PRL_2000,Nguyen_Physica_2001}
and by Monte Carlo simulations
\cite{Wallin_Langmuir_1996_I,Wallin_JPhysChem_1996_II,
Wallin_JPhysChemB_1997_III,Kong_JCP_1998,Jonsson_JCP_2001}.
Nonetheless, much less is known concerning the complexation of a charged
sphere with a like-charged polyelectrolyte. It is only very recently
that we reported in a short communication such a phenomenon in the
strong Coulomb coupling regime \cite{Messina_PRL_2001}.
From a theoretical point of view, the long range Coulomb
interactions of these systems represents a formidable challenge,
and especially the understanding of effective attraction of like charged
bodies has attracted recent attention

In this paper, we elaborate on the complexation between
a sphere and a long flexible polyelectrolyte \textit{both negatively
charged}. While complexation in the strong coupling limit it is expected, we 
report new and rather unexpected chain conformations.
We present MD simulation results without added salt but taking
into account the counterions explicitly. Various Coulomb couplings
as well as different polyelectrolyte linear charge densities are investigated.
A detailed study of the ions (monomer and counterion) distribution
is reported and mechanisms accounting for the different encountered
complex structures are proposed.

The paper is organized as follows. Section \ref{sec.simu} contains
details of our MD simulation model. Section \ref{sec.Strong} is devoted
to the strong Coulomb coupling regime. Section \ref{sec.eps40} is
devoted to the intermediate Coulomb coupling regime. In Sec. \ref{sec.eps80},
we consider the like-charge complexation in the weak Coulomb coupling regime 
corresponding to water solvent.

\section{Simulation method\label{sec.simu}}

The MD method employed here is based on the Langevin equation and
is similar to that employed in previous studies \cite{KG_JCP_1990}.
Consider within the framework of the primitive model one spherical
macroion characterized by a diameter $ d $ and a bare charge \textit{$ Q_{M}=-Z_{M}e $}
(where \textit{e} is the elementary charge and \textit{$ Z_{M}>0 $})
surrounded by an implicit solvent of relative dielectric permittivity
$ \epsilon _{r} $. The polymer chain is made up of $ N_{m} $
monomers of diameter $ \sigma  $. Both ends of the chain are always
charged and each $ 1/f $ monomer is charged so that the chain contains
$ N_{cm}=(N_{m}-1)f+1 $ \textit{charged} monomers. The monomer
charge is $ q_{m}=-Z_{m}e $ (with \textit{$ Z_{m}>0 $}). The
small counterions, assumed all identical, ensure global electroneutrality
and have a diameter $ \sigma  $ and charge \textit{$ +Z_{c}e $}
(with $ Z_{c}>0 $). All these particles making up the
system are confined in an impermeable spherical cell of radius $ R $,
and the spherical macroion is held \textit{fixed} at the center of
the cell. 

The equation of motion of any mobile particle (counterion or monomer)
$i$ reads 

\begin{equation}
\label{eq.Langevin}
m\frac{d^{2}{\mathbf{r}}_{i}}{dt^{2}} =
         -\nabla _{i}U({\mathbf{r}}_{i})-m\gamma \frac{d{\mathbf{r}}_{i}}{dt}+{\mathbf{W}}_{i}(t)\: ,
\end{equation}
%
where \textit{m} is the mass particle (supposed identical for all
mobile species), \textit{U} is the \textit{total} potential force
and $ \gamma  $ is the friction coefficient. Friction and stochastic
force are linked by the dissipation-fluctuation theorem 
$ <{{\mathbf{W}}_{i}}(t)\cdot {{\mathbf{W}}_{j}}(t')>=6m\gamma k_{B}T\delta _{ij}\delta (t-t^{'}) $. 

Excluded volume interactions are introduced via a pure short range
repulsive Lennard-Jones (LJ) potential given by 

\begin{equation}
\label{eq.LJ}
U_{LJ}(r)=\left\{ \begin{array}{l}4\epsilon \left[ \left(
        \frac{\sigma}{r-r_{0}}\right) ^{12} -
        \left( \frac{\sigma }{r-r_{0}}\right) ^{6}\right] +\epsilon ,\\
0,
\end{array}\qquad \right. \begin{array}{l}
\textrm{for}\, \, r-r_{0}<r_{cut},\\
\textrm{for}\, \, r-r_{0}\geq r_{cut},
\end{array}
\end{equation}
%
where $ r_{0}=0 $ for the microion-microion interaction, $ r_{0}=7\sigma  $
for the macroion-microion interaction and $ r_{cut} $ = $ 2^{1/6}\sigma  $
is the cutoff radius. This leads to $ d=2r_{0}+\sigma =15\sigma  $,
whereas the closest center-center distance of the microions to the
spherical macroion is $ a=r_{0}+\sigma =8\sigma  $. The macroion volume fraction
is defined as $ f_{M}=(a/R)^{3} $ and is fixed to $ 8\times 10^{-3} $
with $ R=40\sigma  $.

The pair electrostatic interaction between any pair \textit{ij}, where
\textit{i} and \textit{j} denote either a macroion or a charged microion
(counterion or charged monomer), reads 

\begin{equation}
\label{eq.coulomb}
\frac{U_{coul}(r)}{k_{B}T}=l_{B}\frac{Z_{i}Z_{j}}{r},
\end{equation}
%
where $ l_{B}=e^{2}/4\pi \epsilon _{0}\epsilon _{r}k_{B}T $ is
the Bjerrum length. Energies are measured in units of 
$ \epsilon =k_{B}T $ with $ T=298 $ K. Choosing $ \sigma =3.57 $ \AA\ rrequires
that the Bjerrum length of water at room temperature equals
$ 2\sigma  $ (7.14 \AA). In this work the macroion charge is
fixed at $ Z_{M}=180 $.

The polyelectrolyte chain connectivity is modeled by using a standard
 finitely extensible nonlinear elastic (FENE) potential
in good solvent (see for example Ref. \cite{KG_JCP_1990}), which
reads

\begin{equation}
\label{eq.FENE}
U_{FENE}(r)=\left\{ \begin{array}{l}
-\frac{1}{2}\kappa R^{2}_{0}\ln \left[ 1-\frac{r^{2}}{R_{0}^{2}}\right] ,\\
\infty ,
\end{array}\qquad \right. \begin{array}{l}
\textrm{for}\, \, r<R_{0},\\
\textrm{for}\, \, r\geq R_{0},
\end{array}
\end{equation}
%
where $ \kappa $ is the spring constant-like chosen as $ 1000k_{B}T/\sigma ^{2} $and
$ R_{0}=1.5\sigma  $. These values lead to an equilibrium bond
length $ l=0.8\sigma  $. 

Typical simulation parameters are summarized in Table \ref{tab.simu-param}.
The simulation runs are reported in Table \ref{tab.Runs}. Each simulation
run requires about $ 10^{7} $ MD steps, and equilibrium is typically
reached after $ 5\times 10^{5} $ up to $ 3\times 10^{6} $ steps.
We normally performed between $ 5\times 10^{6} $ and $ 10^{7} $
MD steps to take measurements. We cover the whole range of strength of Coulomb
electrostatic interaction from the strong coupling limit, which is more
theoretical interest, to the weak coupling limit, which corresponds to an
aqueous solvent. 

\begin{table}[b]
\caption{
General data of the simulation model. 
Note that the temperature is used as energy scale for the simulations.}
\label{tab.simu-param}
\begin{ruledtabular}
\begin{tabular}{cc}
 parameters&
\\
\hline
$ \sigma =3.57 $ \AA\ &
 Lennard Jones length units\\
 $ T=298K $&
 room temperature\\
 $ \epsilon =k_{B}T $&
 Lennard Jones energy units\\
 $ Z_{M}=180 $&
 macroion valence\\
$ Z_{m} $&
monomer valence\\
 $ Z_{c} $&
 counterion valence\\
$ N_{c} $&
total number of counterions\\
 $ l_{B} $&
 Bjerrum length\\
 $ a=8\sigma  $&
 macroion-counterion distance of closest approach \\
$ R=40\sigma  $&
simulation cell radius\\
 $ f_{M}=8\times 10^{-3} $&
 macroion volume fraction\\
$ \kappa =1000kT/\sigma ^{2} $&
FENE spring constant\\
$ R_{0}=1.5\sigma  $&
FENE cutoff\\
$ l=0.8\sigma  $&
average bond length\\
$ N_{m} $&
total number of monomers\\
$ N_{cm} $&
 number of charged monomers \\
$ f $&
monomer charge fraction\\
$ \lambda _{PE}=Z_{m}ef/l $&
polyelectrolyte linear charge density \\
\end{tabular}
\end{ruledtabular}
\end{table}

\begin{table}
\caption
{
Specification of the simulated systems. 
The chain radii of gyration $R_g^{(bulk)}$ and $R_g^{(comp)}$ 
are given for an isolated chain (i.e., in the absence of the  colloid) and 
for the complexation case (i.e., in the presence of the colloid) respectively.
Lengths are in  units of $\sigma$.
}
\label{tab.Runs}
\begin{ruledtabular}
\begin{tabular}{cccccccccc}
 parameter&
$ 1/f $ &
$ N_{m} $ &
$ N_{cm} $ &
$ N_{c} $ &
$ Z_{m} $ &
$ Z_{c} $ &
$ l_{B} $ &
$ R_g^{(bulk)} $ &
$ R_g^{(comp)} $ \\
\hline
run $A$&
 1&
 256 &
 256& 
 346&
 2&
 2&
 10&
 3.81&
 6.42\\
run $B$ &
 2 &
 257 &
 129 &
 219&
 2&
 2&
 10&
 3.90&
 8.69\\
run $C$ &
 3 &
 256 &
 86 &
 176&
 2&
 2&
 10&
 3.95&
 8.75\\
run $D$ &
 5&
 256 &
 52&
 142&
 2&
 2&
 10&
 4.49&
 8.86\\
run $E$ &
 1&
 256&
 256&
 346&
 2&
 2&
 4&
 4.40&
 4.8\\
run $F$ &
 2&
 257&
 129 &
 219&
 2&
 2&
 4&
 4.8&
 9.4\\
run $G$ &
 3&
 256&
 86 &
 176&
 2&
 2&
 4&
 5.4&
 9.3\\
run $H$ &
 1&
 256&
 256& 
 346&
 2&
 2&
 2&
 6.6&
 6.1\\
run $I$ &
 2&
 257&
 129 &
 219&
 2&
 2&
 2&
 12&
 13\\
run $J$ &
 3&
 256&
 86 &
 176&
 2&
 2&
 2&
 -&
 21\\
run $K$ &
 1&
 256&
 256&
 436&
 1&
 1&
 2&
 -&
 -\\
\end{tabular}
\end{ruledtabular}
\end{table}

\section{Strong Coulomb coupling\label{sec.Strong}}

First we look at the like-charge complexation in the
strong Coulomb coupling regime. We choose the relative permittivity
$ \epsilon _{r}=16 $, corresponding to $ l_{B}=10\sigma  $, and
divalent microions ($ Z_{m}=Z_{c}=2 $). Such a set of parameters is of special
theoretical interest to study the influence of strong electrostatic correlations.

\subsection{Single charged object\label{sec.SINGLE}}

In this section we first study  \textit{separately} (i) the
spherical macroion and (ii) the flexible polyelectrolyte in the 
presence of their surrounding neutralizing divalent
counterions separately . 
This provides the reference states for the more complicated situation, 
where both of these two objects interact.

\subsubsection{Spherical macroion\label{sec.Macroion_alone}}

Consider an isolated macroion with its surrounding divalent
counterions. A pertinent parameter to describe the Coulomb coupling
for such highly asymmetric electrolyte solution (macroion and counterions)
is the so-called {}``plasma'' parameter $ \Gamma =Z^{2}_{c}l_{B}/a_{cc} $,
where $ a_{cc} $ (which will be determined below) is the average
distance (triangular lattice parameter in the ground state) between
counterions lying on the macroion surface \cite{Rouzina_JCP_1996}. For
the strong Coulomb coupling considered we find $ \Gamma \approx 13 $
(with $ Z_{M}=180 $), and for finite macroion volume fraction (here
$ f_{M}=8\times 10^{-3} $), all counterions lie in the vicinity
of the macroion surface \cite{Messina_PRL_2000,Messina_PRE_2001,Messina_EPL_2000,Messina_EPJE_2001}. 

To characterize the counterion layer \textit{structure}, we compute
the counterion correlation function $ g(s) $ on the surface of
the sphere \cite{Messina_PRE_2001,Messina_EPJE_2001}, defined as:

\begin{equation}
\label{eq.g_r}
c^{2}g(s)=\left\langle \sum _{i\neq j}\delta (s'-s_{i})\delta (s''-s_{j})\right\rangle ,
\end{equation}
%
where $ c=N_{c}/4\pi a^{2} $ is the surface counterion concentration ($
N_{c}=Z_{M}/Z_{c} $ being the number of counterions), and $s=|s'-s''|$ corresponds
to the \textit{arc length} on the sphere of radius $ a $ (center-distance of
closest approach of macroion and counterion).  Each counterion (located in the
vicinity of the surface) is radially projected on the ("contact") shell around
the macroion center of radius $ a=8\sigma $.  Correlation functions are
computed by averaging $g(s)$
over 1000 independent equilibrium configurations which are statistically
uncorrelated. The pair distribution $ g(s) $ is normalized as follows

\begin{equation}
\label{eq.g_r-normalization}
c\int _{0}^{\pi a}2\pi sg(s)ds=(N_{c}-1).
\end{equation}

\begin{figure}
\includegraphics[width = 8.0 cm]{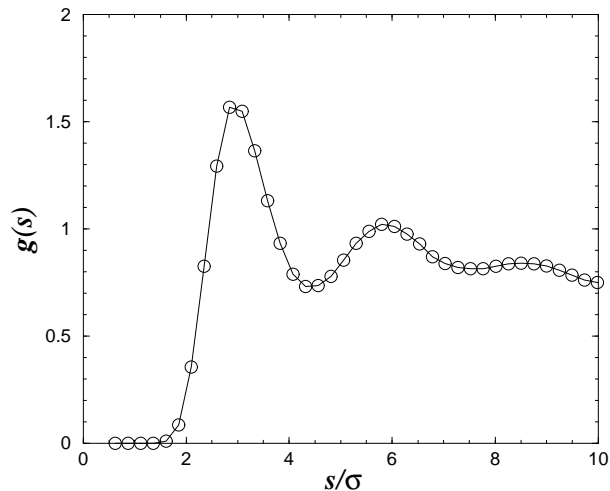}
\caption{\label{fig.gr_macroion}Counterion \textit{surface} correlation function
\protect$ g(s) $ for the 90 divalent counterions ($ Z_{M}=180 $)
in the case of an \textit{isolated} macroion in strong Coulomb coupling
($ l_{B}=10\sigma  $). See Fig. \ref{fig.macroion_alone_snap_eps16}
for a typical equilibrium snapshot.}
\end{figure}

\begin{figure}
\includegraphics[width = 7.0 cm]{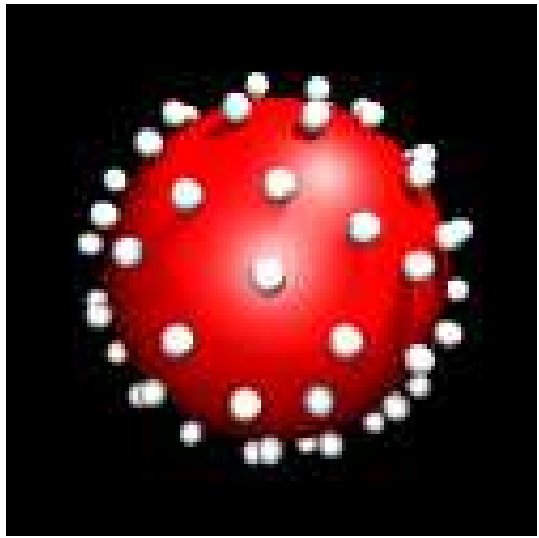}
\caption{\label{fig.macroion_alone_snap_eps16}Snapshot of the equilibrium
counterion structure of an \textit{isolated} macroion with a bare
bare charge $ Z_{M}=180 $ {[}see Fig. \ref{fig.gr_macroion}
for the corresponding $ g(s) ${]}.}
\end{figure}

Because of the \textit{finite} size and the topology of the sphere, $ g(s) $
has a cut-off at $ \pi a $ (=25.1 $ \sigma $) and a \textit{zero} value there.
Therefore, at large values of $s$, $ g(s) $ cannot directly be compared to the
correlation function in an infinite plane.

Results are depicted in Fig. \ref{fig.gr_macroion} for $ Z_{M}=180 $.
The first peak%
\footnote{It is this value $ a_{cc} \approx 3\sigma $ which gives
$ \Gamma \approx 13 $.} 
appears at about $ s= a_{cc} \approx 3\sigma $ whereas the second
peak about $ s \approx 6\sigma  $ and finally the third small peak
around $ s \approx 9\sigma  $. This structure, which is highly
correlated, is referred to as a strongly correlated liquid (SCL) 
\cite{Messina_PRE_2001,Shklowskii_PRE_1999b}, but not yet a Wigner crystal.
A typical equilibrium configuration is depicted in Fig. \ref{fig.macroion_alone_snap_eps16}
where one can see the local arrangement close to a triangular lattice.

\subsubsection{Polyelectrolyte chain \label{sec.PE_alone}}

Now we investigate an isolated polyelectrolyte chain together with its
surrounding divalent counterions in
bulk confined to the same spherical cell of radius $R=40\sigma$.
We consider four monomer charge fractions (i. e. four linear charge
densities) $ f= 1$, 1/2, 1/3 and 1/5.
The chain is made up of $N_{m}=256$  monomers ($N_{m}=257$ for  $f=1/2$). 
while $ Z_{m}=Z_{c}=2 $,
The polymer chain parameters are identical to those of the
complexation case (see Table \ref{tab.Runs}).
The chain extension is characterized by its radius of gyration $R_{g}$
given by

\begin{equation}
\label{eq.Rg}
R_{g}^{2}=\frac{1}{N_{m}}\left\langle \sum ^{N_{m}}_{i=1}({\mathbf{r}}_{i}-{\mathbf{r}}_{CM})^{2}\right\rangle ,
\end{equation}
%
where ${\mathbf{r}}_{CM}$ is the center of mass position of
the chain. 

\begin{figure}[t]
\includegraphics[width = 7.0 cm]{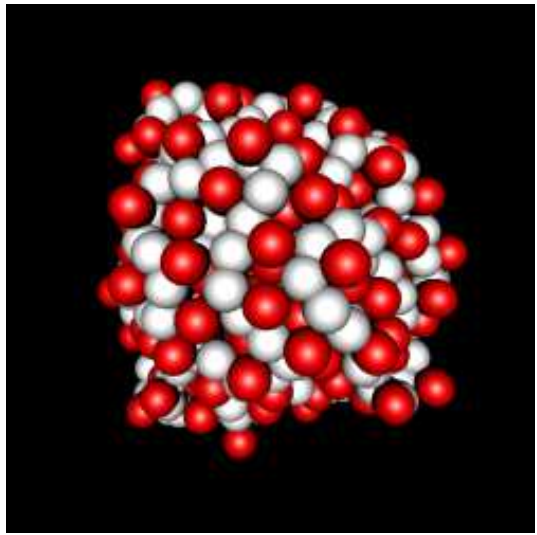}
\caption{\label{fig.PE_alone_snap_eps16}Snapshot of the equilibrium conformation
of an \textit{isolated} polyelectrolyte chain made up of 256 monomers
where all monomers are charged ($ f=1 $). 
Monomers are in white and counterions in dark grey.}
\end{figure}

The corresponding values of $R_g$ can be found in Table \ref{tab.Runs} (see
$R_g^{(bulk)}$ for runs \textit{A-D}).  The chain extension varies little with
$f$ and is roughly given by $ R_{g}\approx 4\sigma $.  For this strong Coulomb
coupling, \textit{all} counterions are {}``condensed'' into the
polyelectrolyte globule for all three linear charge densities considered.  The
strong counterion condensation induces a collapse of the chain, which is by
now well understood
\cite{Stevens_PRL_1993,Stevens_JCP_1995,Winkler_PRL_1997,Brilliantov_PRL_1998,Golestanian_PRL_1999}.
A typical equilibrium chain conformation is shown in Fig.
\ref{fig.PE_alone_snap_eps16} for $ f=1 $. As expected the structure is very
compact and highly ordered. Very similar structures are obtained for $f= 1/2$,
1/3 and 1/5.


\subsection{Complexation \label{sec.Complexation_eps16}}

We now investigate the complexation of a highly charged colloid with
a long flexible polyelectrolyte, both negatively charged. 
Four different parameter combinations,
denoted by run \textit{A}, \textit{B}, \textit{C} and \textit{D},
were investigated which are summarized in Table \ref{tab.Runs}.
Going from run \textit{A} to \textit{D} the polyelectrolyte
charge fraction $f$ decreases from 1 to 1/5. The contour
length of the chain is much larger ($ N_{m}l/d\approx 14 $ times)
than the colloidal particle diameter.

\subsubsection{Observation of the complexation}

\begin{figure}[t]
\includegraphics[width = 8.0 cm]{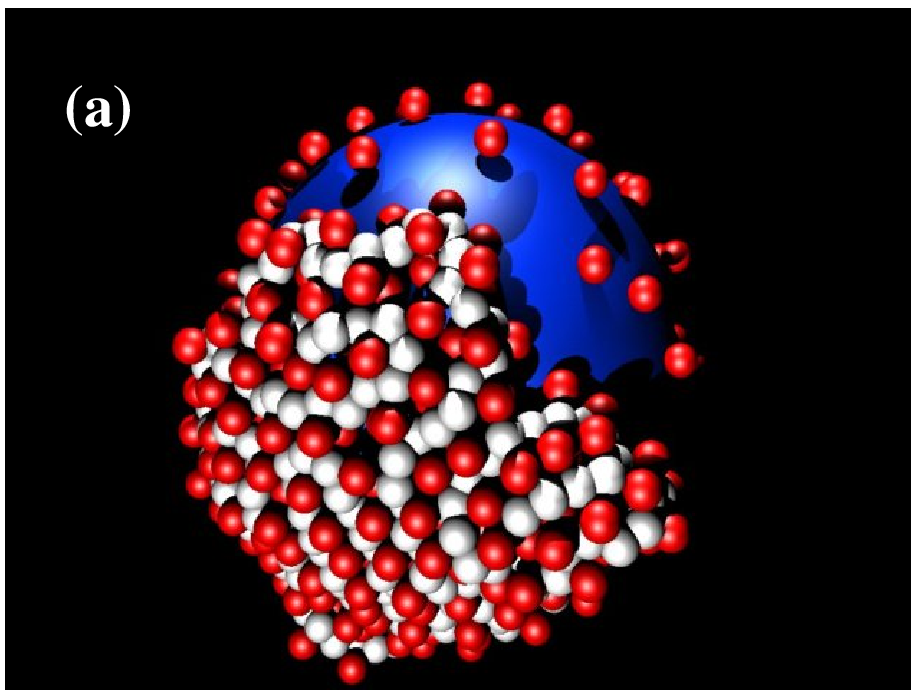}
\includegraphics[width = 8.0 cm]{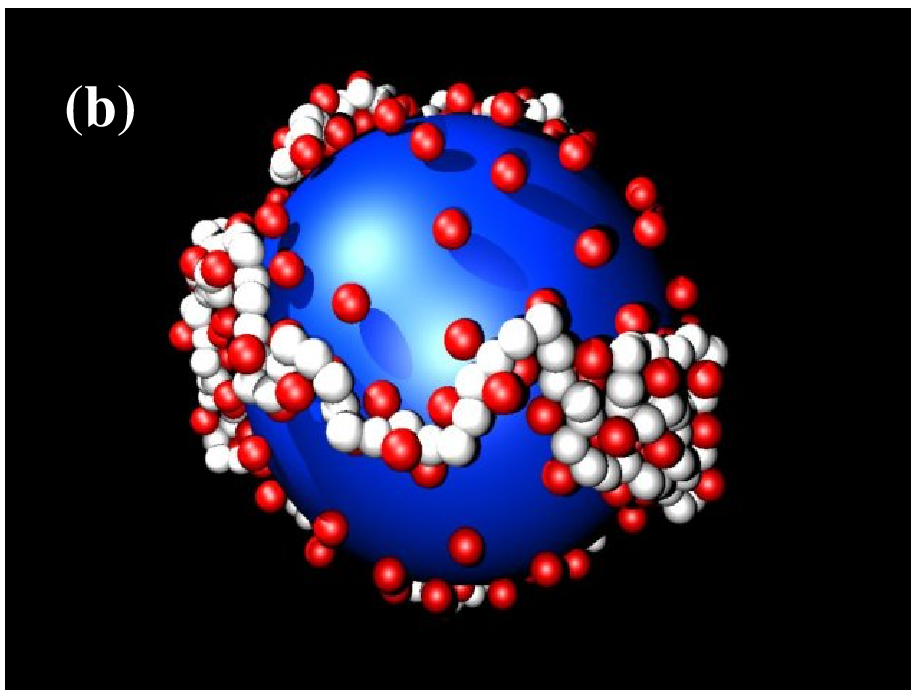}
\includegraphics[width = 8.0 cm]{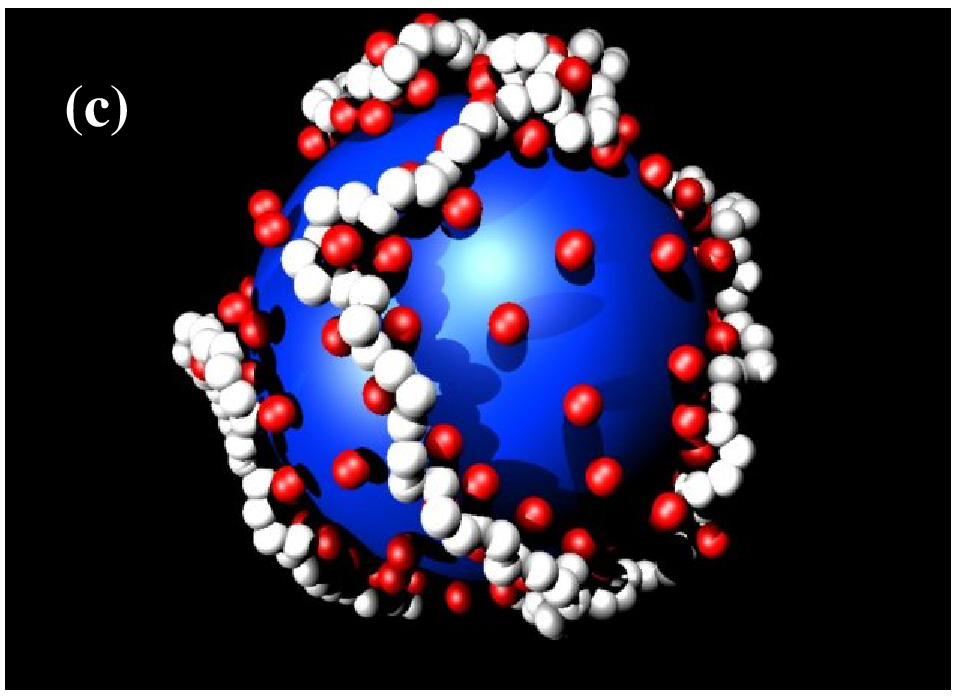}
\includegraphics[width = 8.0 cm]{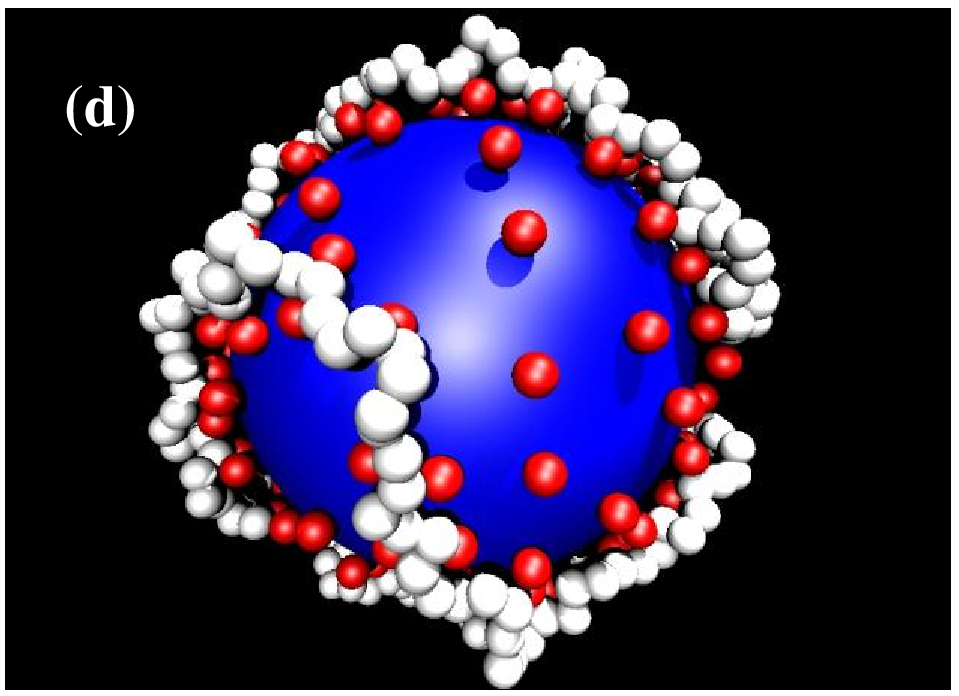}
\caption{\label{fig.complex_snaps_eps16}Typical equilibrium configurations
of the colloid-polyelectrolyte complex in strong Coulomb coupling
($ l_{B}=10\sigma  $) for (a) run $A$ ($ f=1 $),
(b) run $B$ ($ f=1/2 $), (c) run $C$ ($ f=1/3 $) and (d) run $D$
($ f=1/5 $). Monomers are in white and counterions
in red.}
\end{figure}

Figure \ref{fig.complex_snaps_eps16} shows typical equilibrium configurations
of the colloid-polyelectrolyte complex. We notice that in all reported cases
complexation occurs and the polyelectrolyte is completely adsorbed onto the
colloidal surface, that is, in presence of a highly charged colloid, the
polyelectrolyte conformation becomes quasi \textit{two-dimensional} in
contrast to the bulk case (compare Fig. \ref{fig.complex_snaps_eps16} with
Fig. \ref{fig.PE_alone_snap_eps16}).  However the structure of these resulting
complexes depends strongly on the value of $ f $.  For the fully charged
polyelectrolyte case {[}run $A$ with $f=1$ - see Fig.
\ref{fig.complex_snaps_eps16}(a){]} the monomers are closely packed forming a
\textit{two-dimensional} compact Hamiltonian-walk with the condensed
counterions on the polyelectrolyte.  This structure consists of closed packed
lines made from either counterions or monomers.  When the linear change
density is reduced [see Fig. \ref{fig.complex_snaps_eps16}(b-d)], the complex
structures are qualitatively different. In these cases the monomers are no
longer closely packed.  For run \textit{B} {[}$ f=1/2 $, Fig.
\ref{fig.complex_snaps_eps16}(b){]}, the monomers spread more over the
particle surface and the polymer partially wraps around the sphere exhibiting
a quasi two-dimensional surface pearl-necklace structure. For run $C$ {[}$
f=1/3 $, Fig. \ref{fig.complex_snaps_eps16}(c){]} and run $D$ {[}$ f=1/5 $,
Fig. \ref{fig.complex_snaps_eps16}(d){]}, the monomers spread entirely over
the particle surface, and the chain wraps the colloidal particle leading to an
almost isotropic distribution of the monomers around the spherical macroion.

The like-charge complex formation is due to the strong counterion mediated
correlations which are known to induce attractions in the strong Coulomb
coupling regime. Basically, the charged species will try to order locally in a
way which is compatible with the chain connectivity and the macroion surface
constraints.  We now quantify those observations and propose a simple
mechanism to explain the observed conformations.

\subsubsection{Adsorption profile\label{sec.Adsorption-eps16}}

To quantify the adsorption of the monomers and counterions on the macroion
particle surface, we analyze three quantities: (i) the ion radial distribution
function $ n_i(r) $, (ii) the ion fraction $ P_i(r) $ and (iii) the net fluid
charge $ Q_i(r) $ (omitting the macroion bare charge $ Z_{M} $), where $ r $
is the distance from the spherical macroion center. The radial ion
distribution function $ n_{i}(r) $ is normalized as follows

 \begin{equation}
\label{eq.nr}
\int ^{R}_{r_{0}}n_{i}(r)4\pi r^{2}dr=N_{i}\; \; \mathrm{with}\; \; i=c,m;
\end{equation}
%
where $ N_{i} $ is the total number of ions, and the subscripts $ c $ and $ m
$ stand for counterion and monomer respectively.  The reduced integrated ion
radial density $ P_{i}(r) $ is linked to $ n_{i}(r) $ via

\begin{equation}
\label{eq.Pr}
P_{i}(r)=\frac{\int ^{r}_{r_{0}}n_{i}(r^{'})4\pi r^{2}dr^{'}}{N_{i}}\; \; \mathrm{with}\; \; i=c,m.
\end{equation}
%
Note that for $ f<1 $, neutral and charged monomers are all included in the
quantities $ n_{m}(r) $ and $ P_{m}(r) $. The net fluid charge $ Q(r) $ is
given by

\begin{equation}
\label{eq.Qr}
Q(r)=\int ^{r}_{r_{0}}\left[ Z_{c}n_{c}(r^{'})-Z_{m}n_{m}(r^{'})\right] 4\pi r^{2}dr^{'},
\end{equation}
%
where we choose $ e=1 $. 

\begin{figure}[b]
\includegraphics[width = 8.0 cm]{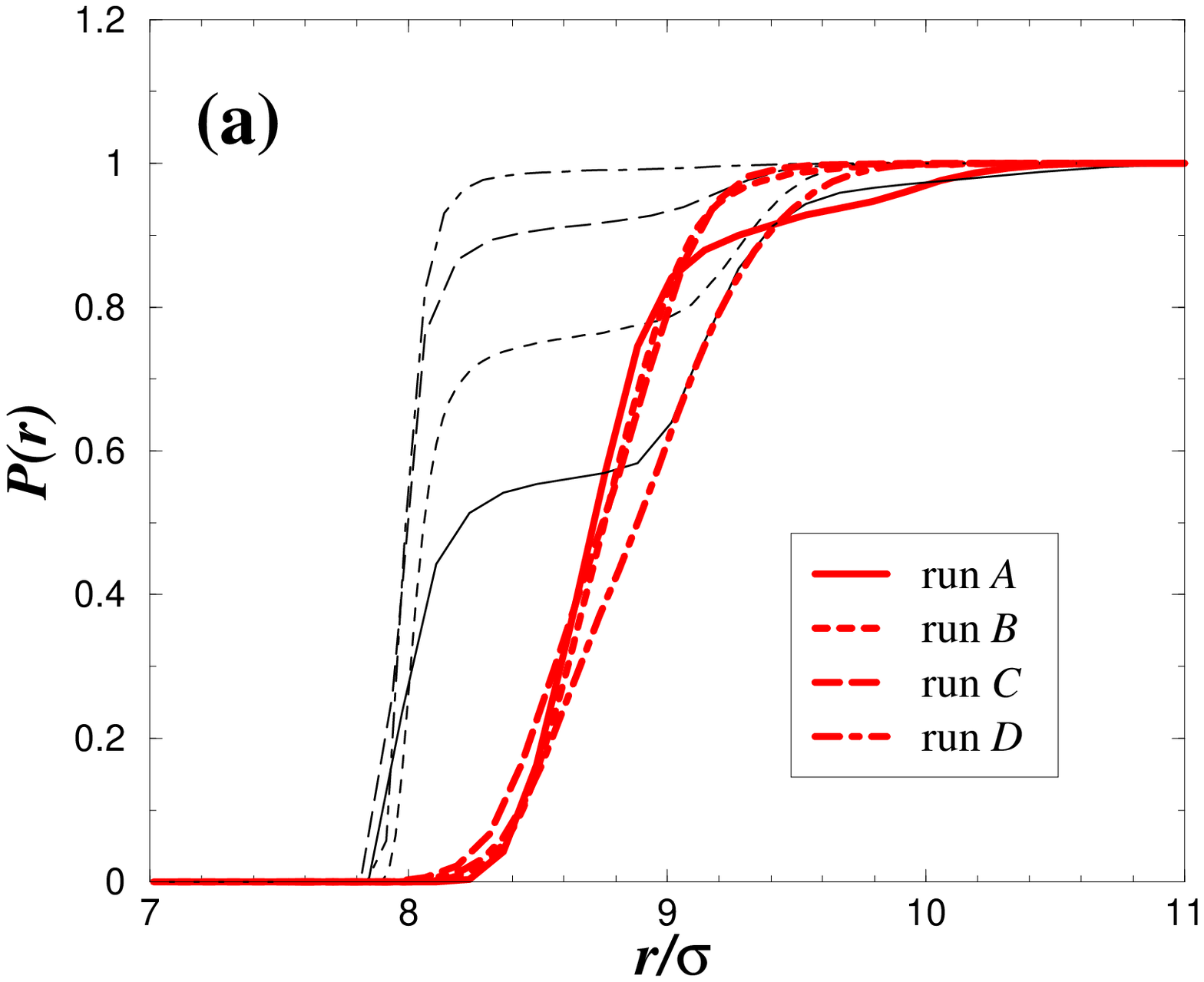}
\includegraphics[width = 8.0 cm]{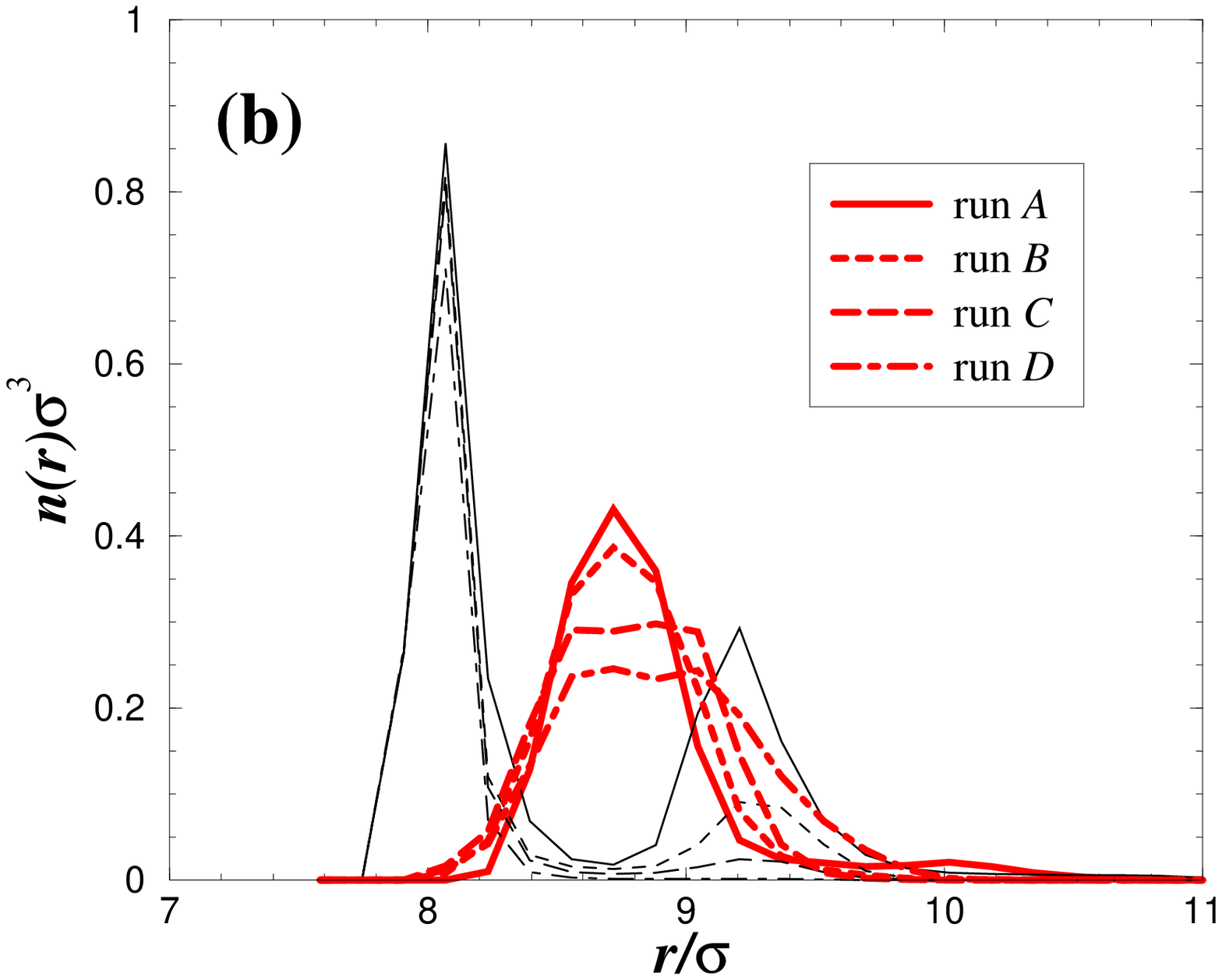}
\includegraphics[width = 8.0 cm]{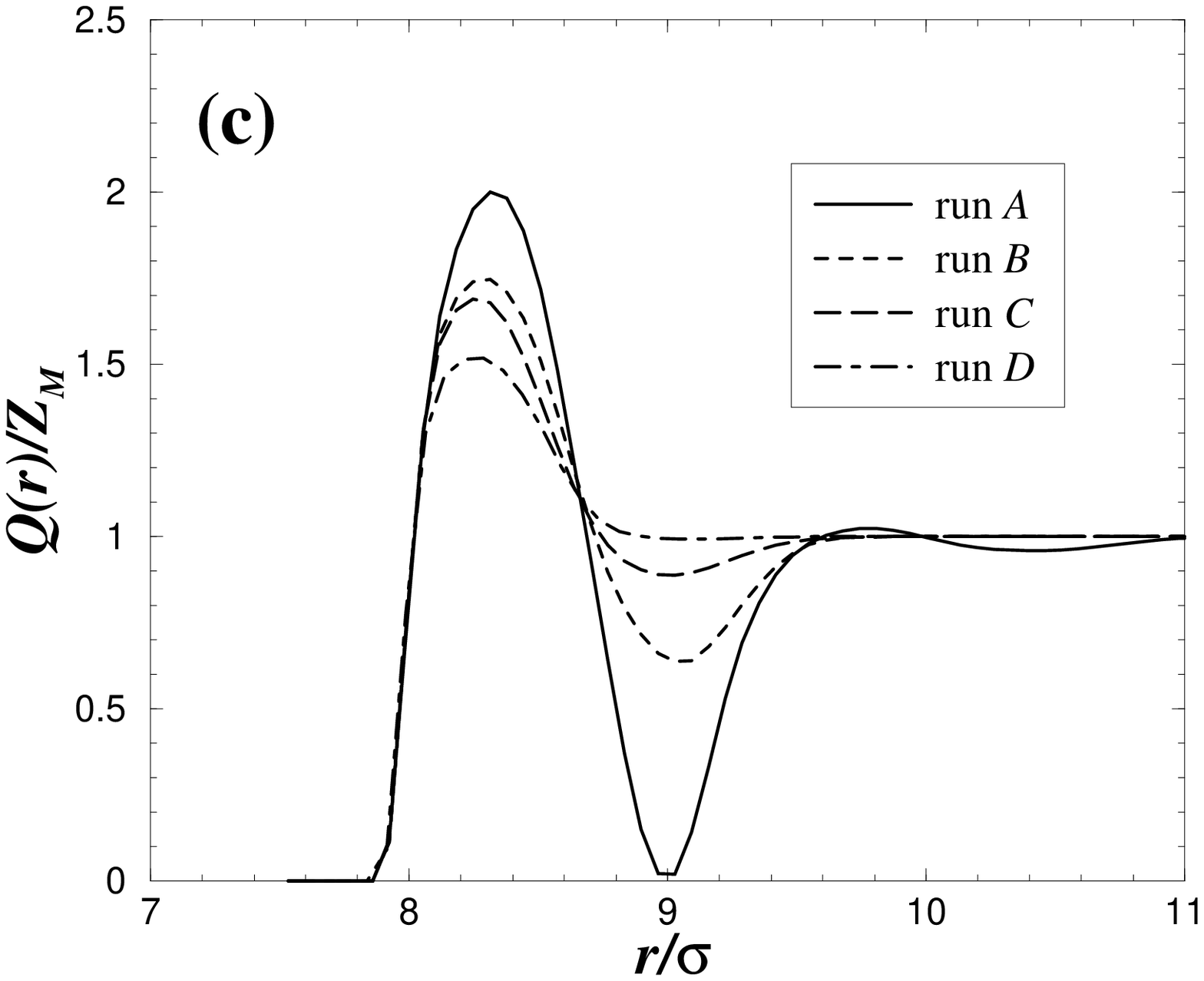}
\caption{Ion adsorption profiles (runs $A-D$) as
a function of the distance $ r $ from the macroion
center. (a) Fraction $ P(r) $ and (b) radial density
$ n(r) $ of counterions (thin lines) and monomers
(thick lines) - (c) reduced net fluid charge $ Q(r)/Z_{M}
$.}
\label{fig.Pr_nr_Qr_eps16}
\end{figure}

Integrated distribution $ P(r) $, radial distribution $ n(r) $ and fluid net
charge $ Q(r) $ profiles are depicted in Figs. \ref{fig.Pr_nr_Qr_eps16}(a-c)
respectively.  Fig. \ref{fig.Pr_nr_Qr_eps16}(a) shows that all atoms for all
runs are condensed within a distance of about $ 10\sigma $ from the colloid
center and more than 80\% of the monomers and counterions are within a
distance of $ 9.3\sigma $ from the colloid center corresponding roughly to two
atomic layers {[}see also the radial distribution in Fig
\ref{fig.Pr_nr_Qr_eps16}(b){]}.  Due to strong electrostatic attraction
between the sphere and the counterions and strong electrostatic repulsion
between the sphere and the charged monomers the first layer ($ r\sim a=8\sigma
$) is exclusively made up of counterions {[}see Figs
\ref{fig.Pr_nr_Qr_eps16}(a-b){]}.  Note that the monomer depletion in this
first counterion layer also concerns \textit{neutral} monomers (runs
\textit{B}-\textit{D}) and this effect is attributed to the chain
connectivity.  This means that the repulsion stemming from the
\textit{charged} monomers impose the polymer structure as long as $f$ and the
Coulomb coupling are sufficiently high, which is the case in the present
study. The height of the first peak in the counterion $ n_{c}(r) $ profile is
almost independent on $ f $. The second ion layer is mixed of monomers and
counterions, however with a majority of monomers {[}see Figs
\ref{fig.Pr_nr_Qr_eps16}(a-b){]}.  Indeed the first monomer peak in the $
n_{m}(r) $ profile (located at $ r\approx 8.7\sigma $) and the second
counterion peak in the $ n_{c}(r) $ profile (located at $ r\approx 9.2\sigma
$) are only separated by roughly $ \Delta r\approx 0.5\sigma $ {[}see Fig
\ref{fig.Pr_nr_Qr_eps16}(b){]}. This leads to a medium position located at $
r\approx 9\sigma $ corresponding to a bilayer thickness.  The height of the
second peak in the counterion $ n_{c}(r) $ profile increases with increasing $
f $ {[}see Fig \ref{fig.Pr_nr_Qr_eps16}(b){]}.

For $ f=1 $ (run $A$), we observe a massive macroion \textit{charge inversion}
of more than 100\% {[}i.e. $ Q(r)/Z_{M}>2 ${]} in the first layer as well as a
strong charge oscillation {[}see Fig. \ref{fig.Pr_nr_Qr_eps16}(c){]}.  Upon
reducing $ f $ (runs $B-D$) the macroion charge overcompensation decreases as
well as the charge oscillation amplitude {[}see Fig.
\ref{fig.Pr_nr_Qr_eps16}(c){]}. This is due to the
fact that upon reducing $f$ less counterions are present and their
correlations change.

\subsubsection{Polyelectrolyte chain radius of gyration}

\begin{figure}[b]
\includegraphics[width = 8.0 cm]{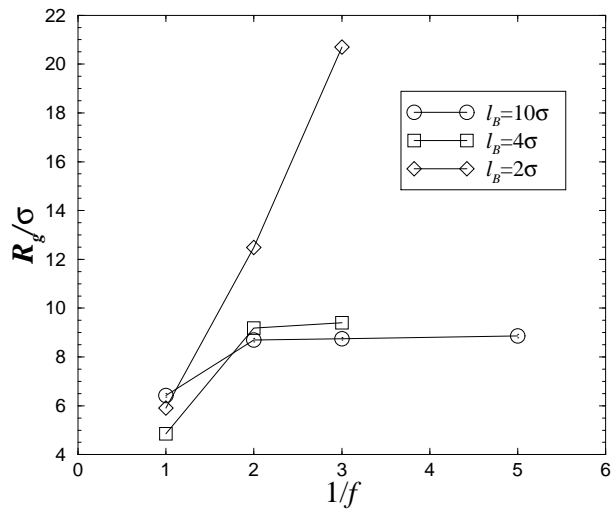}
\caption{Radius of gyration of the polymer as a function
of the polyelectrolyte charge fraction $ f $ for $l_B=10\sigma$ 
(runs \textit{A-D}), $l_B=4\sigma$ (runs \textit{E-F}) 
and $l_B=2\sigma$ (runs \textit{G-H}). }
\label{fig.Rg_complex}
\end{figure}

Next, we investigate the radius of gyration $ R_{g} $ of the chain, i.e. Eq.
(\ref{eq.Rg}), in order to gain insight of the spreading of the monomers over
the sphere. The results reported in Fig. \ref{fig.Rg_complex} (for
$l_B=10\sigma$) show that $ R_{g} $ increases with decreasing decreasing $f$
which demonstrates that the spreading of the monomers over the macroion
surface is enhanced by decreasing the polyelectrolyte charge density.  The
jump in $ R_{g} $ is particularly large between $f=1$ and $f<1$.  This is in
agreement with the visual inspection of the chain conformations presented in
Fig. \ref{fig.complex_snaps_eps16}.  Moreover, the isotropic case (monomers
fully spread over the particle)
corresponding to $ R_{g}\approx 8.7\sigma  $ (%
\footnote{This value corresponds to the the first monomer peak position in the
$ n_{m}(r) $ profile {[}see Fig.\ref{fig.Pr_nr_Qr_eps16} (b){]}.
}) is already reached for $ f=1/2 $ (run \textit{B}). 

\subsubsection{Surface counterion correlation function\label{sec.complex_gr_free}}

\begin{figure}[t]
\includegraphics[width = 7.0 cm]{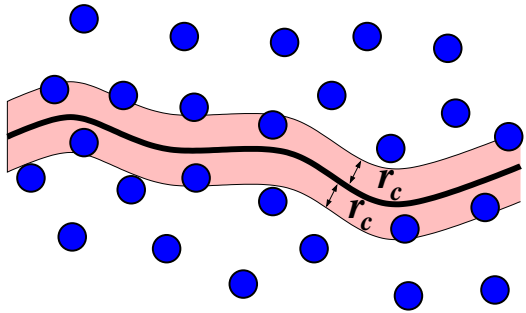}
\caption{
Schematic view of the chain intercepting
counterions within a distance $r_{c}$. 
}
\label{fig.ghost_chain}
\end{figure}

In this section we are interested in determining the structure of the
{}``free'' counterions which are \textit{not condensed} onto the
polyelectrolyte chain. Counterions are called ``condensed'' on the
polyelectrolyte chain when they lie within a distance $ r_{c}=1.2\sigma $
perpendicular to the chain (Fig. \ref{fig.ghost_chain}).  All other
counterions are called ``free'', although they are still adsorbed onto the
colloidal surface. To characterize the structure of the free counterions we
proceed in the same way as in Sec.  \ref{sec.Macroion_alone}.  The surface
free counterion correlation function $ g_{free}(s) $ is now given by

\begin{equation}
\label{eq.gr_free}
c_{free}^{2}g_{free}(s) = 
   \left\langle \sum _{i\neq j}\delta (s'-s_{i})\delta (s''-s_{j})\right\rangle ,
\end{equation}
%
where the sum in Eq. (\ref{eq.gr_free}) is restricted to the free counterions, and
$c_{free}=N_{free}/4\pi a^{2}$ is the surface free counterion concentration,
with $N_{free}$ being the average number of free counterions. The
normalization is obtained as follows

\begin{equation}
\label{eq.gr_free_normalization}
c_{free}\int _{0}^{\pi a}2\pi sg_{free}(s)ds=(N_{free}-1).
\end{equation}
%

\begin{figure}[b]
\includegraphics[width = 8.0 cm]{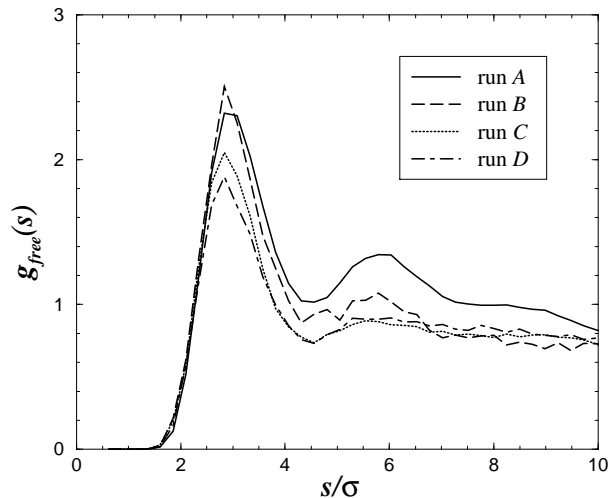}
\caption{Free counterion (unbounded to the
polyelectrolyte) surface correlation function $ g_{free}(s) $
for runs \textit{A-D}.}
\label{fig.gr_complex_free_coun}
\end{figure}

Results are depicted in Fig. \ref{fig.gr_complex_free_coun} for runs
\textit{A-D.} The important result is that the first peak of $ g_{free}(r) $
is located at the same position ($ s\approx 3\sigma  $) as in the
{}``unperturbed'' case of an isolated macroion (without polyelectrolyte)
studied in Sec. \ref{sec.Macroion_alone} (see Fig. \ref{fig.gr_macroion}).
Although the second peak of $ g_{free}(s) $ is less pronounced
than in the {}``unperturbed'' case (compare Fig. \ref{fig.gr_complex_free_coun}
with Fig. \ref{fig.gr_macroion}), the local order of the free counterion
structure is still high as can be visually inspected on the snapshots
sketched in Fig. \ref{fig.complex_snaps_eps16}. 
Thus the adsorbed chain only affects the counterion distribution significantly
in its immediate neighborhood.

\subsubsection{Polyelectrolyte overcharging\label{sec.PE_OC_eps16}}

We now show that the concept of polyelectrolyte \textit{overcharging}
can be used to explain the observed complex structures. 
Let $N_{cd}$ be the number of counterions we consider as condensed
onto the polyelectrolyte.
Then the overcharging ratio $ \chi _{PE} $
is defined as

\begin{equation}
\label{eq.OC-PE}
\chi _{PE}=\frac{N_{cd}}{N_{cm}},
\end{equation}
%
which is merely the ratio between the amount of the \textit{total}
condensed counterion charge and the polyelectrolyte \textit{bare}
charge. 

This ``overcharging'' can also be analytically predicted by 
the simple assumption that the presence of the polyelectrolyte (with its
counterions) does not affect the free counterion distribution%
\footnote{This mainly holds for the first correlational peak which is the most
important for the present discussion, as we showed in the previous 
Sec. \ref{sec.complex_gr_free}.
}. 
Let us consider the
bare charged chain plus its own neutralizing counterions as an uncharged object
that  gets overcharged
by intercepting all counterions (of the macroion) whose center lie
within a ribbon of width $ 2r_{c} $ and area $ A_{rib}=2r_{c}N_{m}l $
(Fig. \ref{fig.ghost_chain}). If $ c $ is the counterion (of
the macroion) concentration then the theoretical overcharging ratio $\chi _{th}$
is merely given by

\begin{equation}
\label{eq.OC-theory-a}
\chi _{th}=1+\frac{A_{rib}c}{N_{cm}}=1+\frac{2r_{c}N_{m}lZ_{M}}{N_{cm}Z_{c}4\pi a^{2}},
\end{equation}
%
and since the number of charged monomers $ N_{cm} $ is given by
$ N_{cm}=(N_{m}-1)f+1 $, Eq. (\ref{eq.OC-theory-a}) reduces for
$ N_{m}\gg 1 $ to

\begin{equation}
\label{eq.OC-theory-b}
\chi _{th}\sim 1+C/f,
\end{equation}
%
with $C=\frac{2r_{c}lZ_{M}}{Z_{c}4\pi a^{2}}$.

\begin{figure}[b]
\includegraphics[width = 8.0 cm]{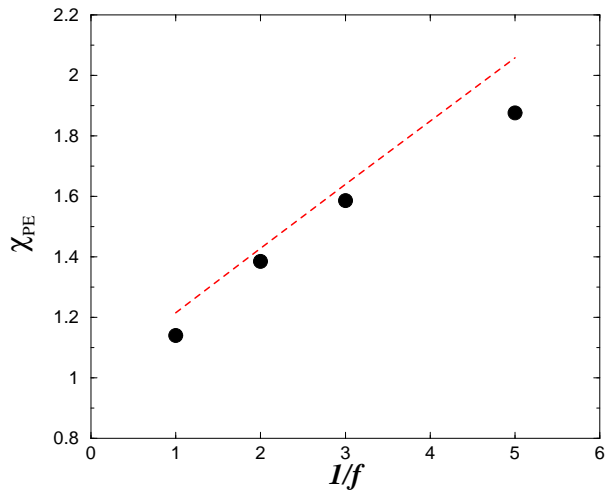}
\caption{
Polyelectrolyte overcharge as a function
of $ f $ (runs \textit{A-D}). The dashed line corresponds
to the theoretical prediction where Eq. (\ref{eq.OC-theory-a}) was
used.
}
\label{fig.OC_PE_eps16}
\end{figure}

Results are presented in Fig. \ref{fig.OC_PE_eps16} and the corresponding
values can be found in Table \ref{tab.OC-eps16}. It indicates that
in all cases overcharging occurs (i.e., $ \chi _{PE}>1 $),
and that it increases with decreasing polyelectrolyte charge density.
We have excellent agreement (less than 10\% difference) between simulation
results and our toy model {[}Eq. (\ref{eq.OC-theory-a}){]}. In turn it explains
why $ \chi _{PE} $ varies almost linearly with $ 1/f $ in our simulations.

\begin{table}[t]
\caption{Polyelectrolyte overcharge $ \chi _{PE} $
values as a function of $ f $ \textit{.}}
\label{tab.OC-eps16}
\begin{ruledtabular}
\begin{tabular}{cccc}
 Run&
$ 1/f $&
$ \chi _{PE} $ - MD&
$ \chi _{PE} $ - Theory\\
\hline
\textit{A}&
1&
 $ 1.140\pm 0.002 $&
 1.21\\
\textit{B} &
2&
$ 1.39\pm 0.02 $&
 1.43\\
\textit{C} &
3&
 $ 1.59\pm 0.02 $&
 1.64\\
\textit{D}&
5&
$ 1.88\pm 0.03 $&
 2.06\\
\end{tabular}
\end{ruledtabular}
\end{table}

The \textit{f}-dependency of the complexation structure can be explained
through the overcharging. For this we consider the
overcharged polyelectrolyte as a \textit{dressed}
(or \textit{renormalized}) chain [bare chain + counterions]
with an (or \textit{renormalized}) linear charge
density $ \lambda ^{*}_{PE}=-(\chi _{PE}-1)\lambda _{PE} $ that obviously has
the opposite sign of $ \lambda _{PE} $ \cite{NOTE_Many_Body}.
Similarly one can define the \textit{renormalized} charge of a  monomer
as 
\begin{equation}
\label{eq.q*}
q^{*}_{m}=-(\chi _{PE}-1)q_{m}.
\end{equation}
%
Using Eq. (\ref{eq.q*}) and the results of Fig. \ref{fig.OC_PE_eps16} this
shows that $ q^{*}_{m} $ increases with increasing $ 1/f $.  The overcharging
leads to an effective local repulsion of the monomers, and subsequently to a
bond stiffening of the chain. This in turn explains why the chain expands with
increasing $ 1/f $ (see Fig. \ref{fig.Rg_complex} and Fig.
\ref{fig.complex_snaps_eps16} for the corresponding structures)
\cite{NOTE_Many_Body_b}.

\section{Intermediate Coulomb coupling \label{sec.eps40}}

In this section we are dealing with a higher dielectric constant ($ \epsilon _{r}=40 $),
meaning that we consider weaker Coulomb coupling ($ l_{B}=4\sigma  $).
Experimentally this could correspond to using alcohol as a solvent. We
consider a set of three runs $E-G$ with $ l_{B}=4\sigma  $ (see
Table \ref{tab.Runs}). Thus these systems are, up to a shorter Bjerrum length 
$ l_{B}$, identical  to runs $A-C$.
It will be helpful to start the discussion with the description of the observed complex
microstructures.

\subsection{Complex microstructure\label{sec.structure_eps40}}

\begin{figure}
\includegraphics[width = 7.0 cm]{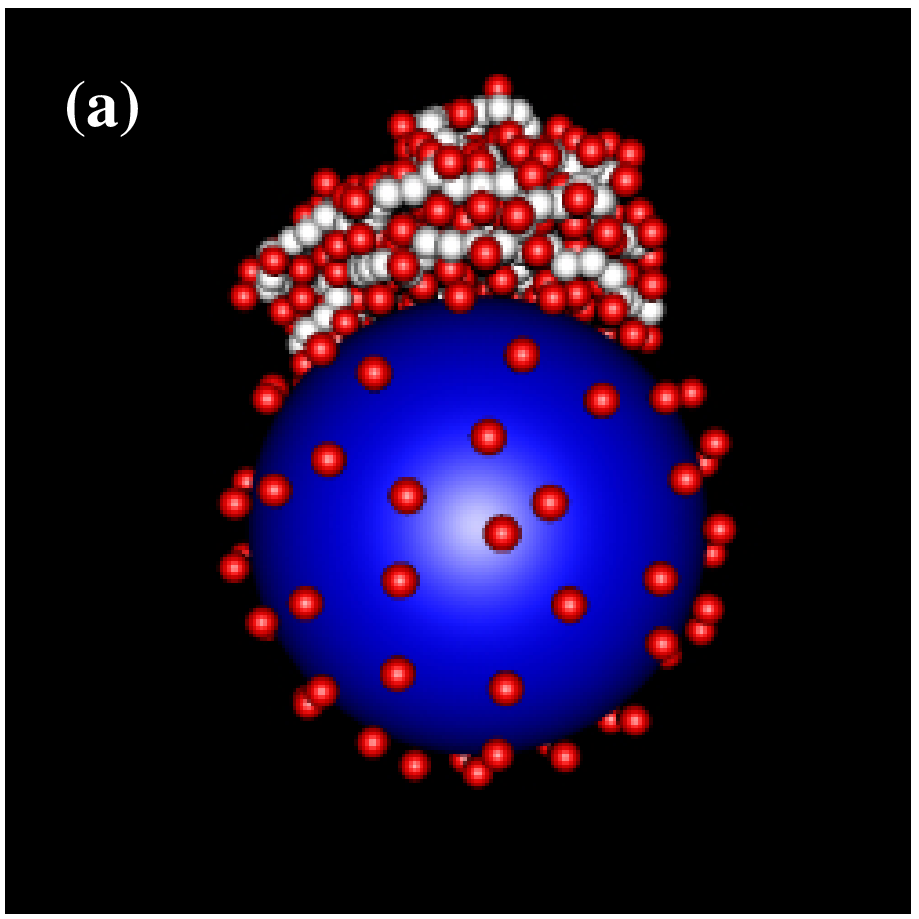} 
\includegraphics[width = 7.0 cm]{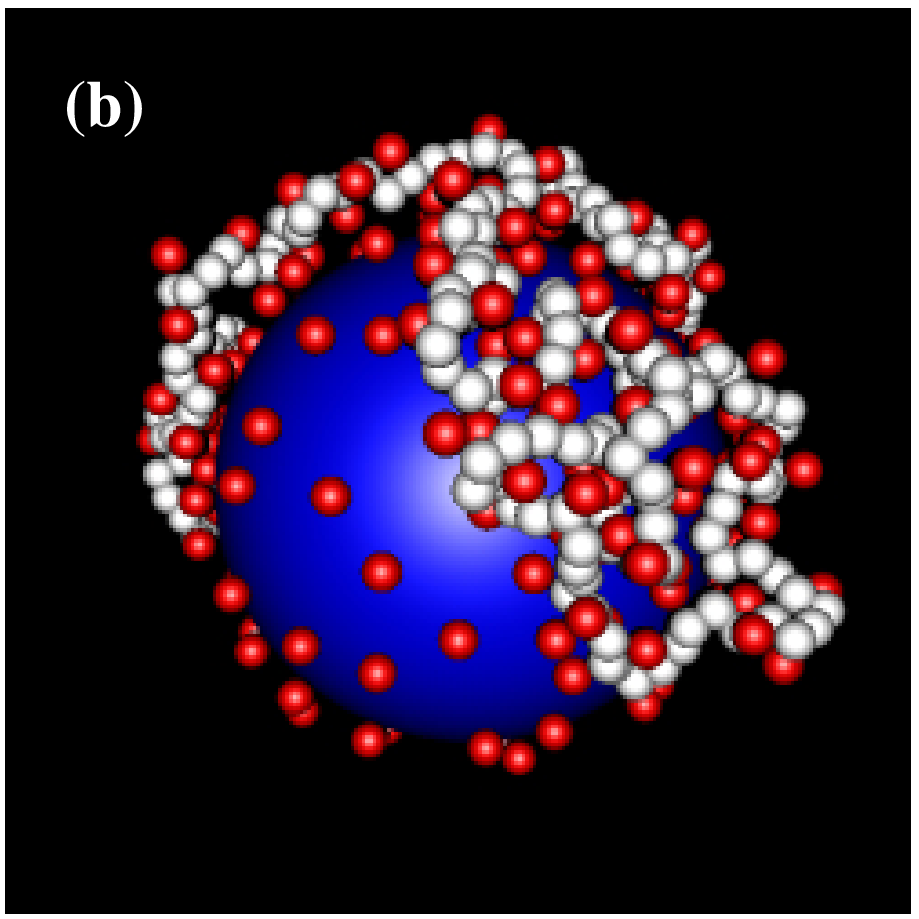} 
\includegraphics[width = 7.0 cm]{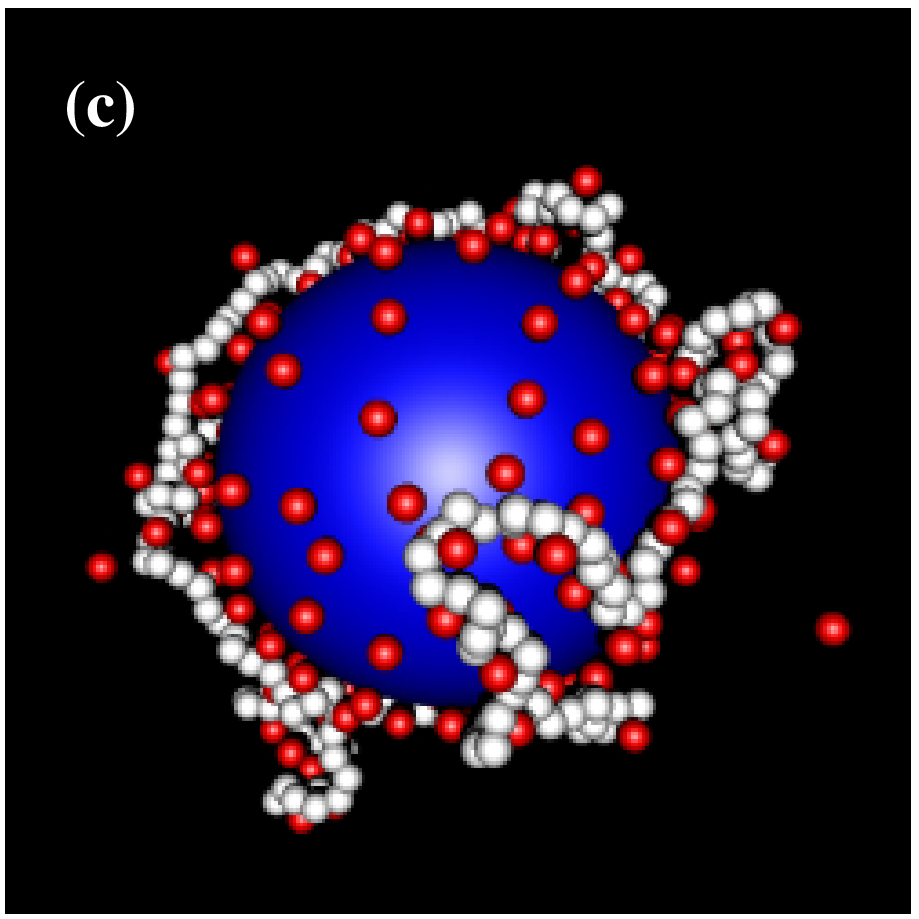} 
\caption{
Typical equilibrium configurations
of the colloid-polyelectrolyte complex in moderate Coulomb coupling
($ l_{B}=4\sigma  $) for (a) run $E$ ($ f=1 $),
(b) run $F$ ($ f=1/2 $) and (c) run $G$ ($ f=1/3 $).
Monomers are in white and counterions in red.
}
\label{fig.complex_snaps_eps40}
\end{figure}

Typical equilibrium macroion-polyelectrolyte complex structures are sketched
in Fig. \ref{fig.complex_snaps_eps40}.  When the polyelectrolyte is fully
charged ($ f=1 $, run $E$), Fig. \ref{fig.complex_snaps_eps40}(a) shows again
a strongly compact chain conformation.  But in the present situation the chain
does not spread on the macroion surface as it was the case in the strong
Coulomb coupling {[}compare Fig.  \ref{fig.complex_snaps_eps40}(a) with Fig.
\ref{fig.complex_snaps_eps16}(a){]}.  In fact the conformation of the charged
chain in presence of the macroion is very similar to the bulk conformation
{[}compare \ref{fig.complex_snaps_eps40}(a) with Fig.
\ref{fig.PE_alone_snap_eps16}{]}.

By reducing the monomer charge fraction, Figs. \ref{fig.complex_snaps_eps40}(b-c)
show that the complex microstructure for runs $F,G$ is again qualitatively
different from the fully charged case (run $E$). For these smaller
polyelectrolyte linear charge densities ($ f=1/2 $ and $ f=1/3 $),
the chain conformation is again almost wrapping around the colloid. For
$ f=1/2 $ (run $F$) the peal-necklace structure observed in the
strong Coulomb coupling {[}see Fig. \ref{fig.complex_snaps_eps16}(b){]}
does not appear here, instead small loops appear 
{[}see Fig. \ref{fig.complex_snaps_eps40}(b){]}.
For the smallest monomer charge fraction ($ f=1/3 $, run $G$) the
monomers fully spread over the macroion surface and the conformation
is not compact, and also small loops appear {[}see Fig.
\ref{fig.complex_snaps_eps40}(c){]}. 
Again it is observed that upon reducing the polymer charge density the chain
expands, but this time it expands also into the radial direction away from
the macroion.

The forthcoming sections are devoted to study in more detail the monomer
and counterion distributions.

\subsection{Adsorption profile}

\begin{figure}[t]    
\includegraphics[width = 8.0 cm]{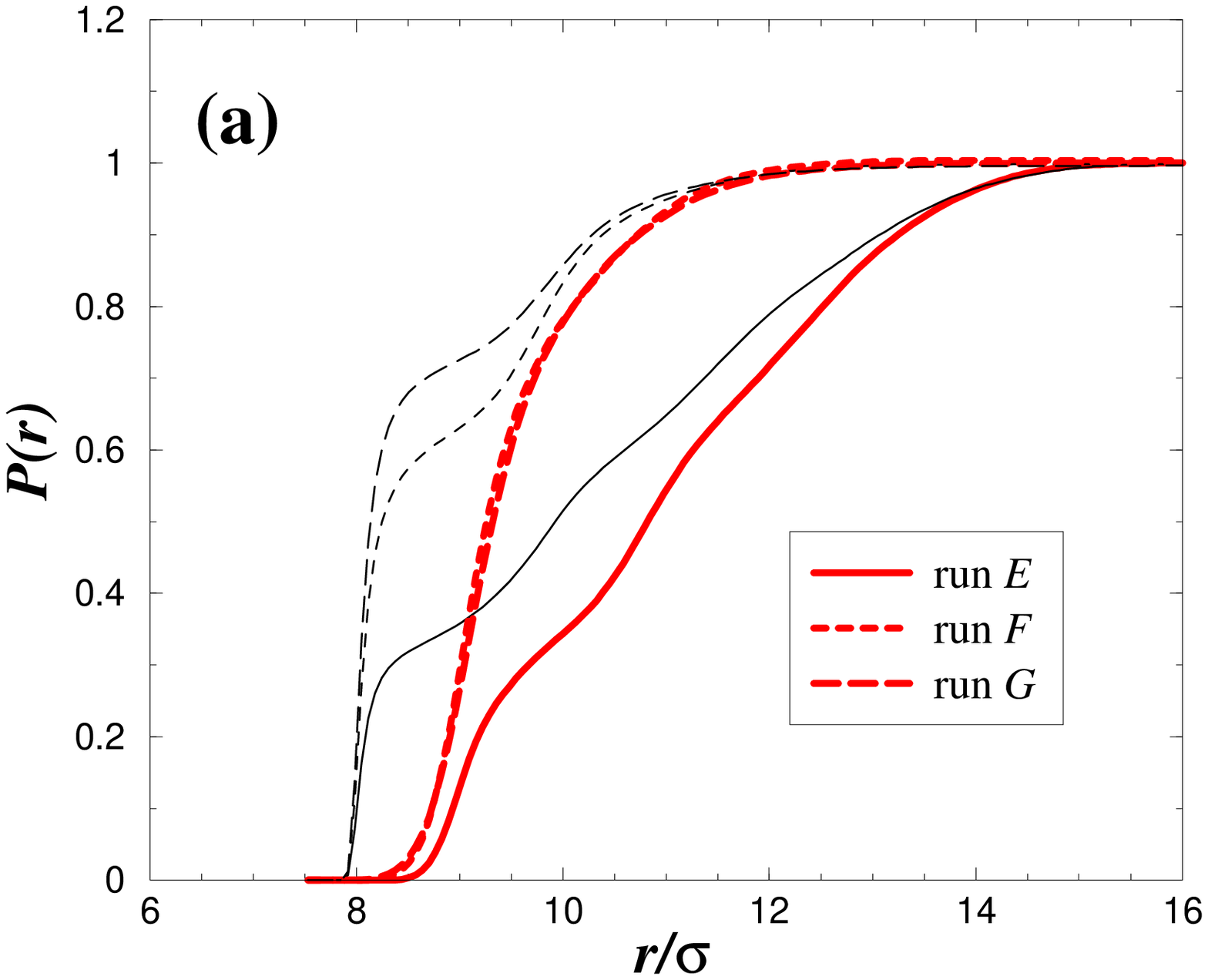}
\includegraphics[width = 8.0 cm]{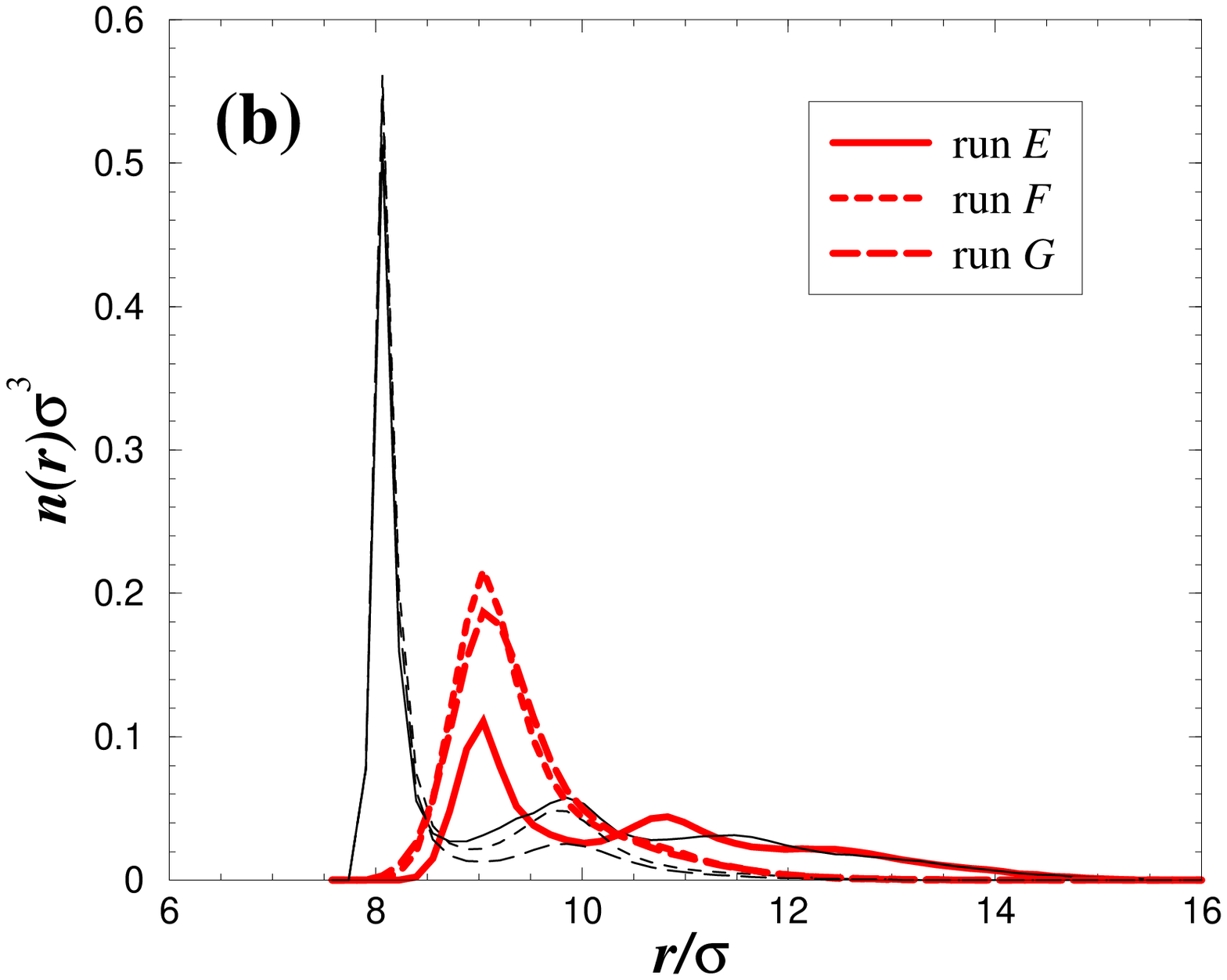}
\includegraphics[width = 8.0 cm]{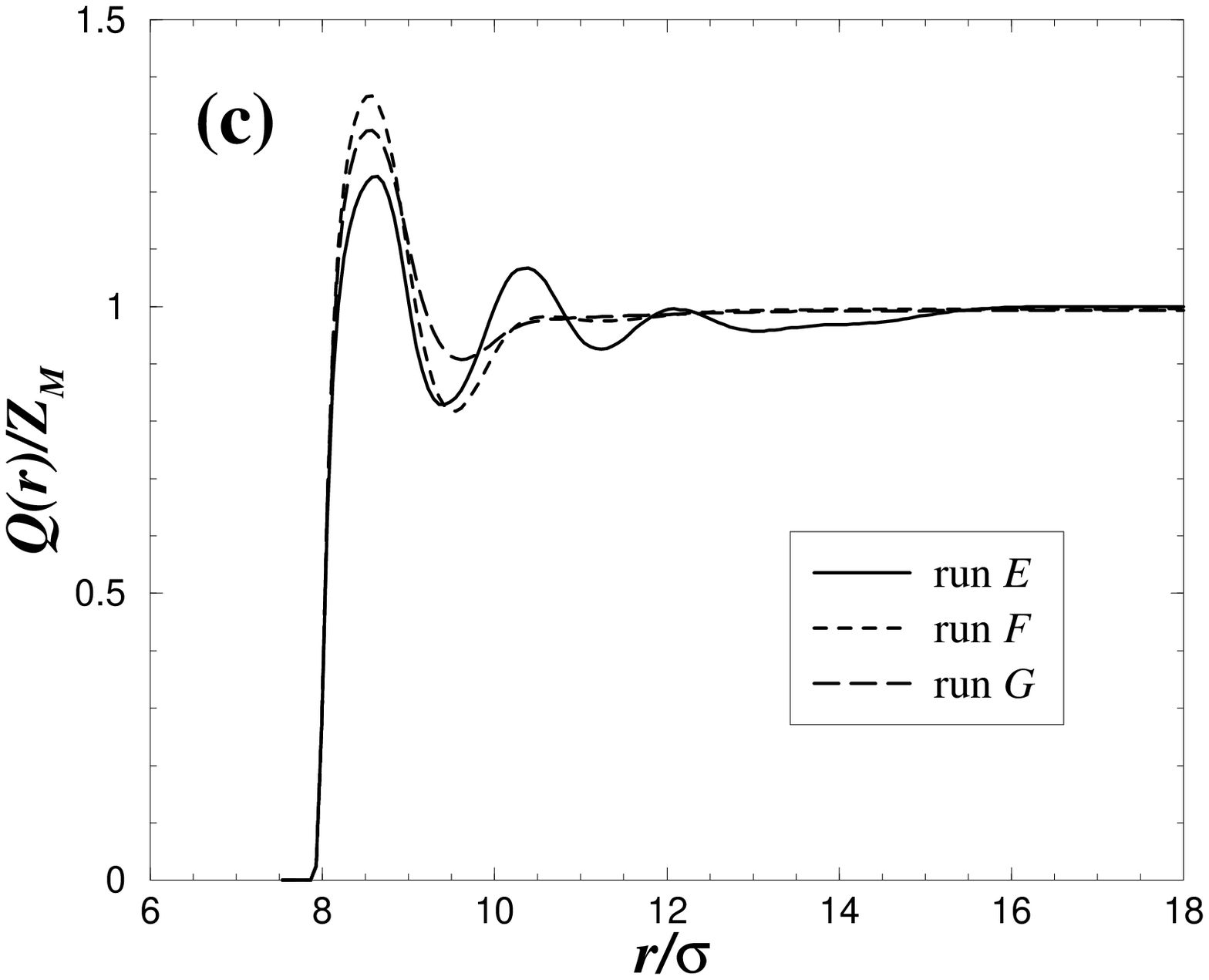}
\caption{
Ion adsorption profiles (runs $E-G$) as
a function of the distance $ r $ from the macroion
center. (a) Fraction $ P(r) $ and (b) radial density
$ n(r) $ of counterions (thin lines) and monomers
(thick lines) - (c) reduced net fluid charge $ Q(r)/Z_{M} $. }
\label{fig.Pr_nr_Qr_eps40}
\end{figure}

Integrated distribution $P(r)$, radial distribution $n(r)$
and fluid net charge $Q(r)$ profiles are given in Figs. \ref{fig.Pr_nr_Qr_eps40}(a-c)
respectively. 
In a general manner, the fraction of monomers and counterions
in the vicinity of the macroion surface is clearly smaller than the
one obtained in the strong Coulomb coupling as expected {[}compare
Fig. \ref{fig.Pr_nr_Qr_eps40}(a) with Fig. \ref{fig.Pr_nr_Qr_eps16}(a){]}.
Also the width in $P(r)$ profile of adsorbed ions
is enlarged with decreasing Coulomb coupling {[}compare Fig. \ref{fig.Pr_nr_Qr_eps40}(a)
with Fig. \ref{fig.Pr_nr_Qr_eps16}(a){]}. These features show that
the monomer adsorption is more diffuse (in the normal direction to the macroion
sphere) as expected for a weaker Coulomb coupling. Concerning the
ion density $ n(r) $ profile {[}see Fig. \ref{fig.Pr_nr_Qr_eps40}(b){]},
it is interesting to note that the height of the \textit{first} peak
in the monomer density $ n_{m}(r) $ profile is twice smaller for
$ f=1 $ than for $ f<1 $, whereas the one from the counterion
density profile is almost independent on $ f $. In parallel, the
macroion charge overcompensation as well as charge oscillation amplitudes
are clearly reduced compared to the strong Coulomb regime {[}compare
Fig. \ref{fig.Pr_nr_Qr_eps40}(c) with Fig. \ref{fig.Pr_nr_Qr_eps16}(c){]}.

For the fully charged polyelectrolyte case ($ f=1 $, run $E$) the
$ n_{m}(r) $ profile in Fig. \ref{fig.Pr_nr_Qr_eps40}(b) shows
a strong second monomer peak and a weaker third one in agreement with
the snapshot of Fig. \ref{fig.complex_snaps_eps40}(a). The radial
monomer ordering naturally goes along with a counterion ordering in
anti-phase. In other words, multilayering of different chain segments 
occurs, but without strong adsorption of the macroion [compare 
Fig. \ref{fig.complex_snaps_eps40}(a) with
Fig. \ref{fig.complex_snaps_eps16}(a){]}. 
Therefore we find three charge oscillations {[}see Fig. \ref{fig.Pr_nr_Qr_eps40}(c){]}
against only two in the strong Coulomb coupling {[}see Fig. \ref{fig.Pr_nr_Qr_eps16}(c){]}. 

For smaller linear charge density, Fig. \ref{fig.Pr_nr_Qr_eps40}(a) and Fig.
\ref{fig.Pr_nr_Qr_eps40} (b) indicate that for runs $F$ ($ f=1/2 $) and $G$ ($
f=1/3 $) the chain is almost fully adsorbed to the macroion surface without
monomer chain multilayering {[}i. e., no appearance of monomer second peak in
the $ n_{m}(r) $ profile - see Fig. \ref{fig.Pr_nr_Qr_eps40}(b){]}.  However,
the conformation is a little bit swollen probably due to the onset of loop
formation, compare the snapshots in Fig. \ref{fig.complex_snaps_eps40}.  As in
the strong Coulomb coupling case, we find here only one charge oscillation for
$ f=1/2 $ and $ f=1/3 $, and the amplitude of the reduced net fluid charge
decreases with decreasing $ f $.

\subsection{Polyelectrolyte chain radius of gyration}

Results for the  radius of gyration $ R_{g} $ of the polymer chain
are reported in Fig. \ref{fig.Rg_complex} (for $l_B=4\sigma$). As for the high
Coulomb coupling case with $ l_{B}=10\sigma  $, $ R_{g} $ increases
with decreasing $ f $. 
However, for $ f=1 $, here we obtain $ R_{g}\approx 4.8\sigma  $
which is clearly smaller than the value $ R_{g}\approx 6.4\sigma  $
obtained in run $A$ (see Fig. \ref{fig.Rg_complex}).
This proves that for $ l_{B}=4\sigma  $ the chain conformation
stays as a globule, since we already found that
for the same chain length and with $ l_{B}=10\sigma  $ the chain
conformation was compact and two-dimensional. In the case under consideration
($ f=1 $, $ l_{B}=4\sigma  $), the value of $ R_{g}\approx 4.8\sigma  $
is almost identical  to the one obtained in the bulk where $ R_{g}= 4.4\sigma  $ 
(see Table \ref{tab.Runs}).

%
%

Upon reducing $f$ (run $F$ and $G$), the chain is much more expanded
and it is found that $ R_{g}\approx 9.4\sigma  $. Taking into account
the fact the chain is still adsorbed for both systems (run $F$ and
$G$) as was found in the analysis of the adsorption profile (see Fig.
\ref{fig.Pr_nr_Qr_eps40}), one deduces that the spreading of the
monomers is very important as soon as $ f<1 $. A comparison with
the strong Coulomb coupling shows that values of $ R_{g} $ for
$ l_{B}=4\sigma  $ are systematically larger than those of $ R_{g} $
with $ l_{B}=10\sigma  $ (see Fig. \ref{fig.Rg_complex} and
Table \ref{tab.Runs}), indicating again that the chain fluctuates
more in the outward macroion radial direction at weaker Coulomb coupling.

\subsection{Polyelectrolyte overcharging}

In order to check if the local polyelectrolyte overcharge is responsible
for the expansion of the chain upon reducing $ f $ as was demonstrated
in the strong Coulomb coupling for $l_{B}=10\sigma$ in Sec.
\ref{sec.PE_OC_eps16}, we again consider the overcharging ratio $\chi _{PE}$
defined by Eq. (\ref{eq.OC-PE}) with the same condensation distance
$r_{c}=1.2\sigma $ as was done for $ l_{B}=10\sigma  $ 
in Sec. \ref{sec.PE_OC_eps16}.

Numerical values of $ \chi _{PE} $ are can be found in Table \ref{tab.OC_PE_eps40}.
It clearly shows that polyelectrolyte overcharge is negligible (i.
e., $ \chi _{PE}\approx 1 $) for the present Coulomb coupling regime
whatever the value of $ f $. Consequently one can not explain the
expansion of the chain with increasing $ 1/f $ with a polyelectrolyte
overcharge mechanism. We will give clear and qualitative arguments
in Sec. \ref{sec.Discussion} that account for these conformations.

\begin{table}[t]
\caption{
 Polyelectrolyte overcharge $ \chi _{PE} $
values as a function of $ f $ (runs $E-G$).
}
\label{tab.OC_PE_eps40}
\begin{ruledtabular}
\begin{tabular}{ccc}
Run&
$ 1/f $&
$ \chi _{PE} $\\
\hline
$E$&
1&
$ 1.043\pm 0.002 $\\
$F$&
2&
$ 1.089\pm 0.003 $\\
$G$&
3&
$ 1.065\pm 0.006 $\\
\end{tabular}
\end{ruledtabular}
\end{table}

\section{Weak Coulomb coupling \label{sec.eps80}}

This part is devoted to aqueous solutions where the Bjerrum length is $
l_{B}=2\sigma =7.14$ \AA\ ccorresponding to the dielectric constant $ \epsilon
_{r}\approx 80 $ of \textit{water}.  Such systems will be referred to as the
weak Coulomb coupling regime.  We have considered a set of three runs $H-J$
(see Table \ref{tab.Runs}) identical to the previous ones but with a shorter
Bjerrum length $ l_{B}=2\sigma $.  Again it is helpful to start with the
description of the observed complex microstructures.

\subsection{Complex microstructure\label{sec.structure_eps80}}

\begin{figure}
\includegraphics[width = 7.0 cm]{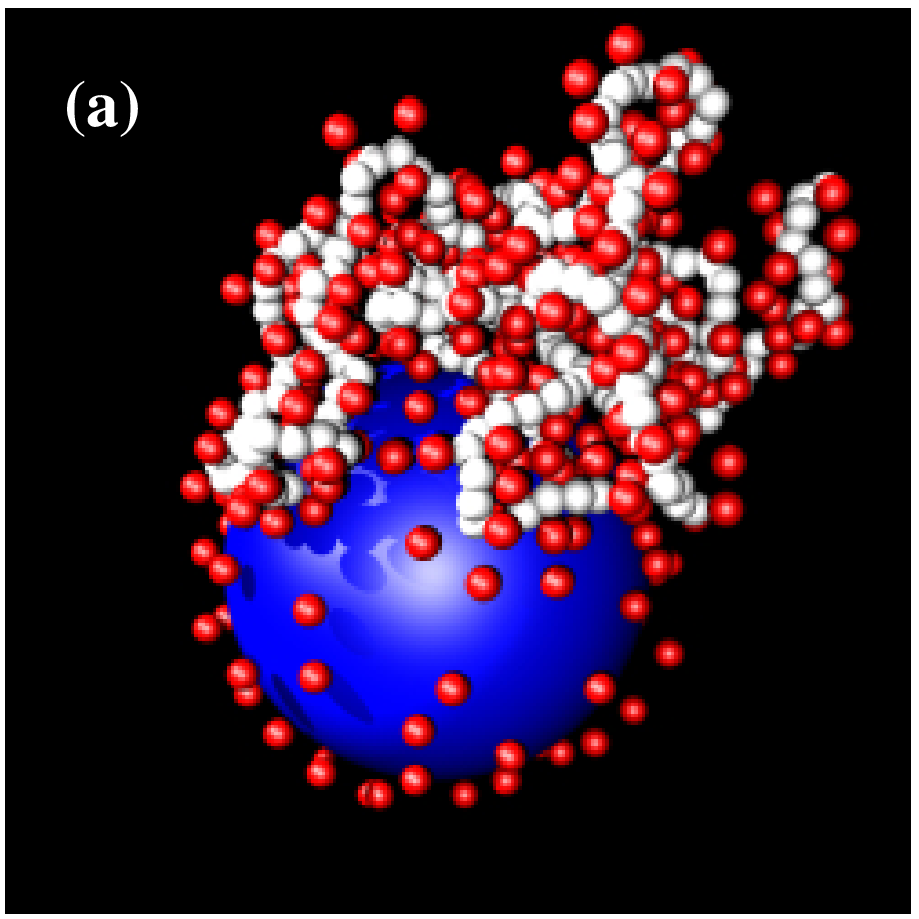}
\includegraphics[width = 7.0 cm]{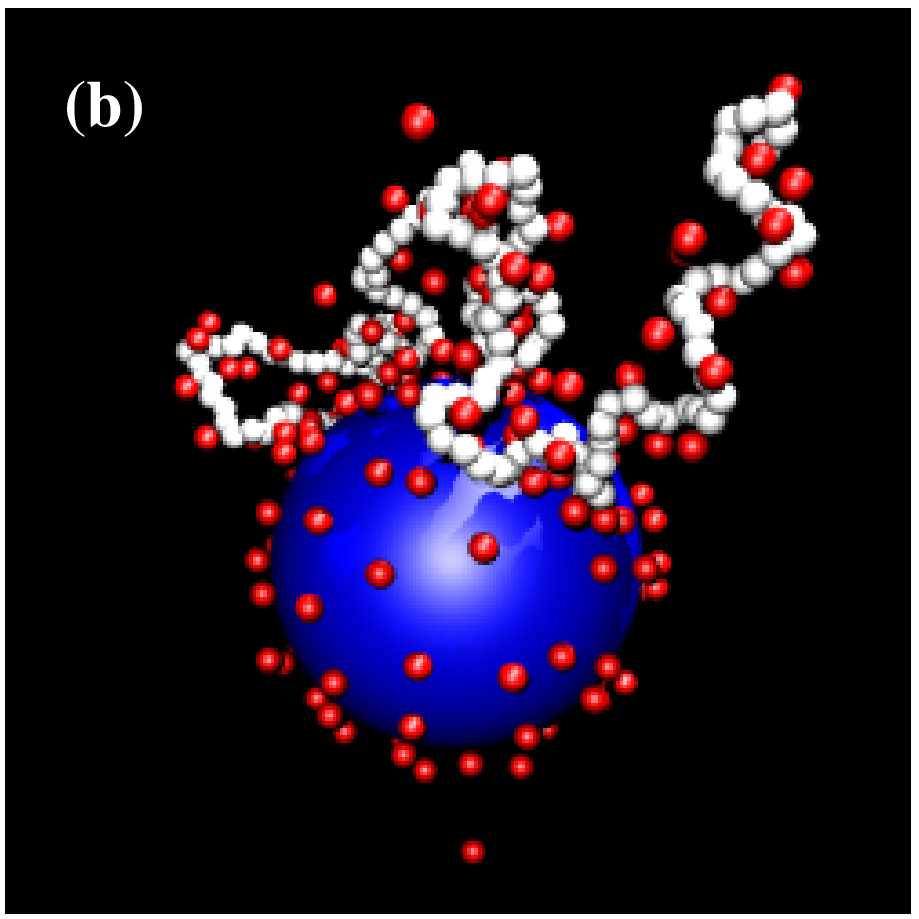}
\includegraphics[width = 7.0 cm]{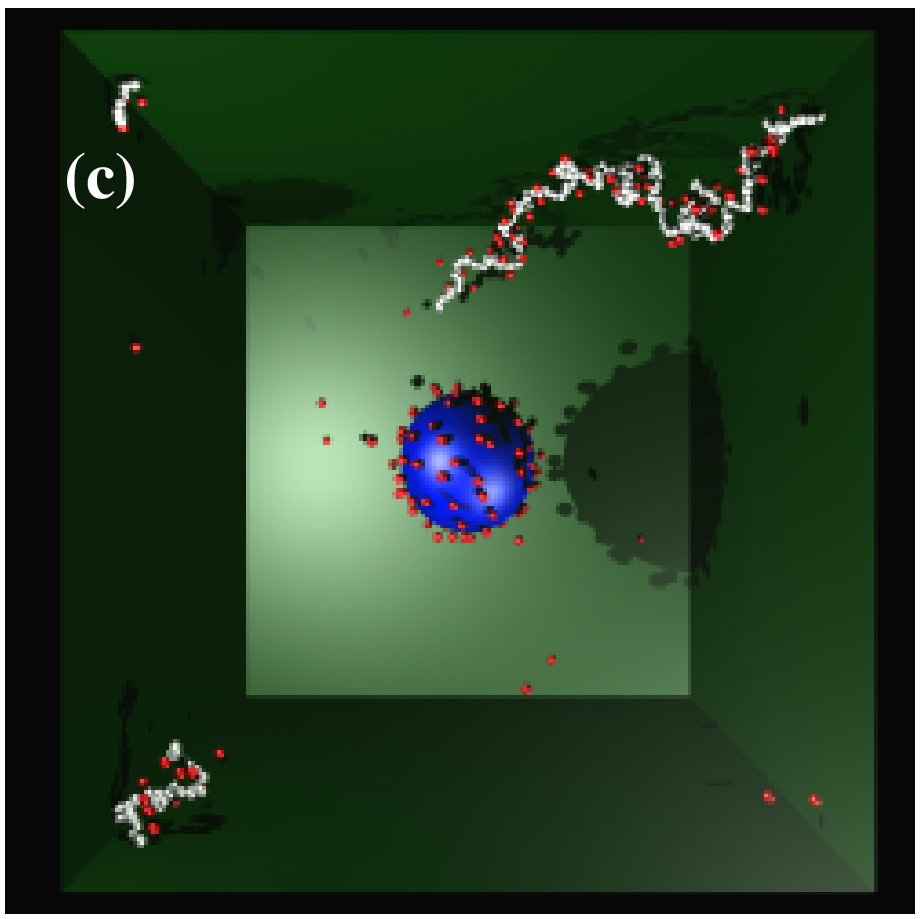}
\caption{
  Typical equilibrium configurations of the colloid-polyelectrolyte complex in
  weak Coulomb coupling ($ l_{B}=2\sigma $) for (a) run $H$ ($ f=1 $), (b) run
  $I$ ($ f=1/2 $) and (c) run $J$ ($ f=1/3 $). Monomers are in white and
  counterions in red.  Snapshot (c) was obtained with periodic boundary
  conditions.  }
\label{fig.complex_snaps_eps80}
\end{figure}

Typical equilibrium macroion-polyelectrolyte complex structures can be found
in Fig. \ref{fig.complex_snaps_eps80}.  For all investigated cases (runs
$H-J$), one finds that the polymer \textit{never} adopts a ``two-dimensional''
conformation. A comparison of the bulk value of $R_g$ and the $R_g$ of the
chain in the complexed situation (compare Table \ref{tab.Runs}) reveals that
the chain conformation is only weakly affected by the macroion. For the fully
charged polymer ($ f=1 $, run $H$), the conformation is again rather compact
but without exhibiting a strong monomer-counterion ordering (within the
polymeric aggregate) as it was the case for higher Coulomb coupling regimes
{[}compare Fig.  \ref{fig.complex_snaps_eps80}(a) with Fig.
\ref{fig.complex_snaps_eps40}(a) and Fig. \ref{fig.complex_snaps_eps16}(a){]}.
However we do have an effective macroion-polyelectrolyte attraction, and the
dense monomer-counterion aggregate is adsorbed onto the colloidal surface.

For $ f=1/2 $ (run $I$), the chain conformation is more expanded than for
\textit{$ f=1 $} {[}compare Fig. \ref{fig.complex_snaps_eps80}(b) with Fig.
\ref{fig.complex_snaps_eps80}(a){]}\textit{.} Nevertheless we do have polymer
adsorption with the formation of chain loops. Therefore even for couplings
which are typical of aqueous systems, our simulations show that like-charge
complexation can occur for $ f=1 $ and $ f=1/2 $ with divalent ions ($
Z_{m}=Z_{c}=2 $).

For even lower linear charge density ($ f=1/3 $, run $J$), Fig.
\ref{fig.complex_snaps_eps80}(c) shows polymer desorption from $ f=1/2 $ to $
f=1/3 $.  In the same time there is a certain {}``\textit{counterion
  release}'' for $ f=1/3 $, meaning that not all counterions are in the
vicinity of the highly charged objects (macroion and polyelectrolyte).  We
carefully checked that, with periodic boundary conditions, the same features
qualitatively appear, namely chain desorption from $ f=1/2 $ to $ f=1/3 $.

\subsection{Adsorption profile}

\begin{figure}
\includegraphics[width = 8.0 cm]{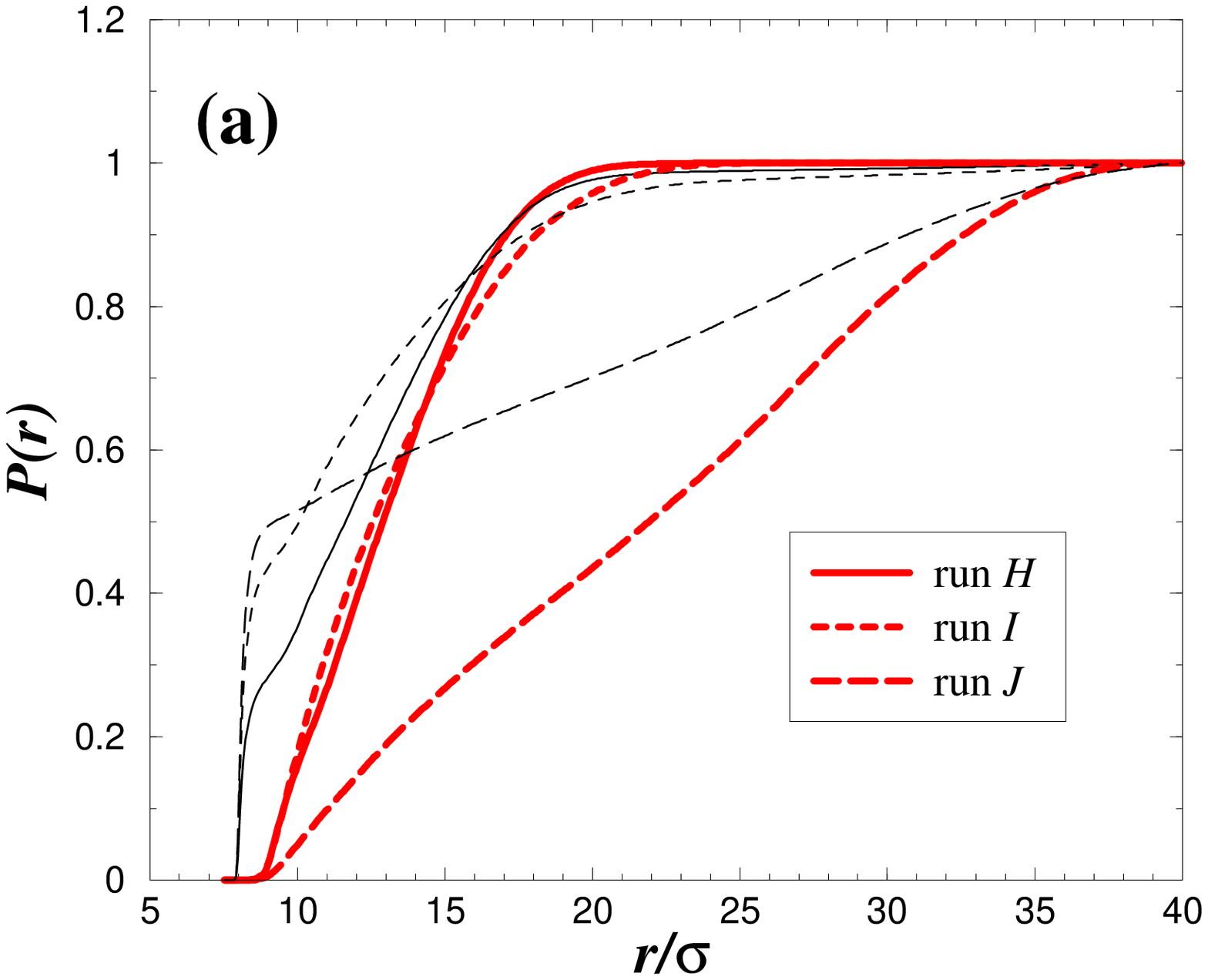}
\includegraphics[width = 8.0 cm]{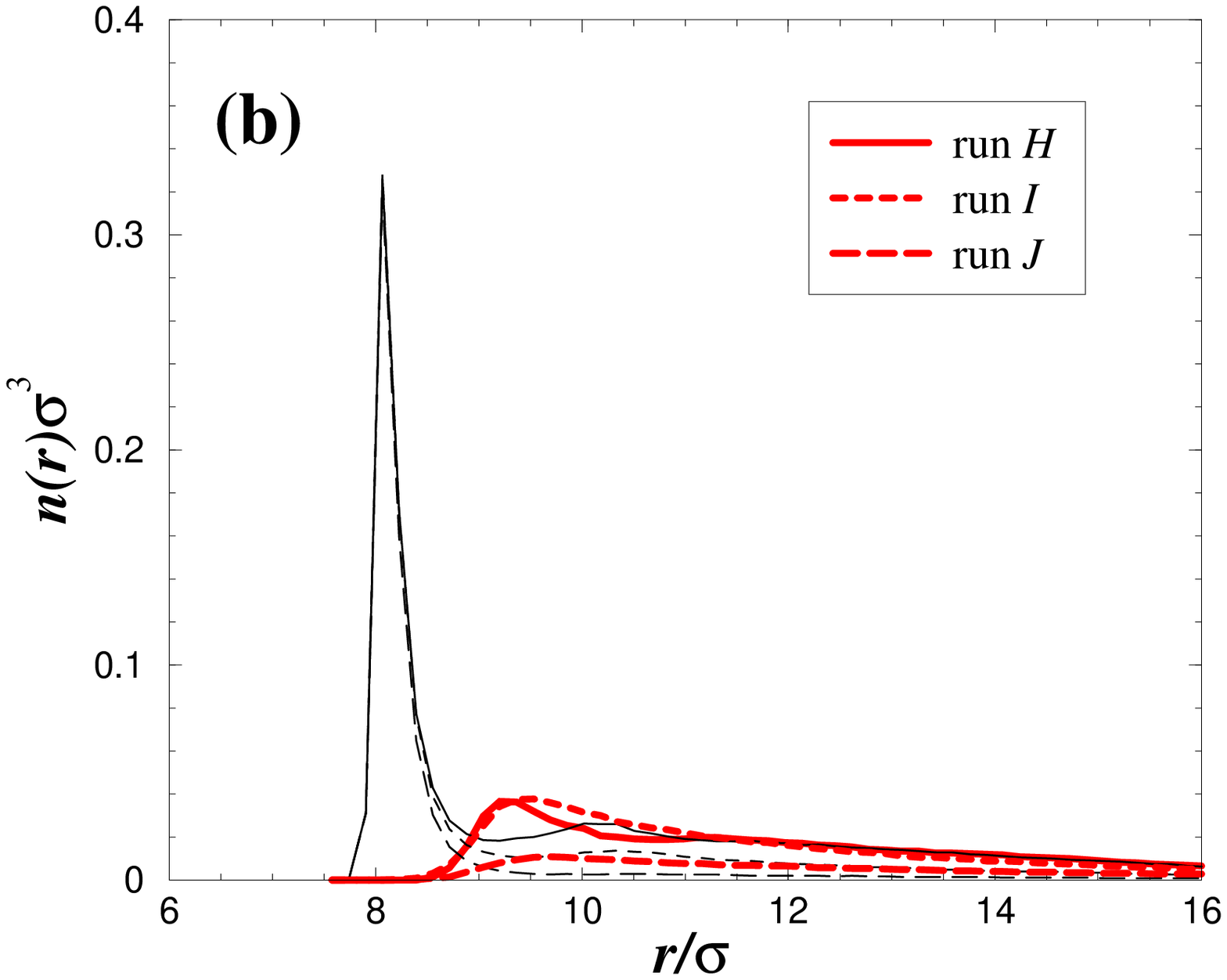}
\includegraphics[width = 8.0 cm]{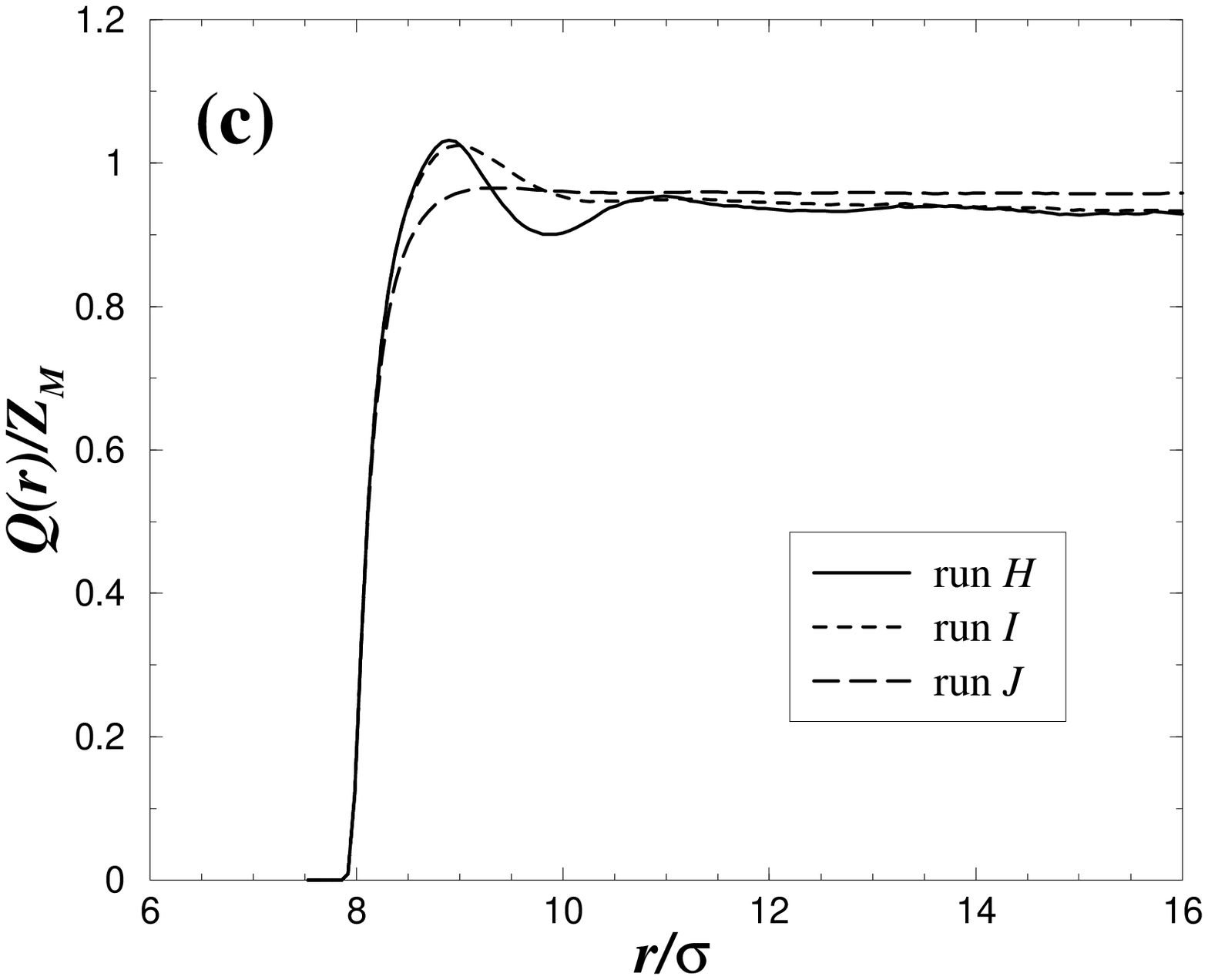}
\caption{
  Ion adsorption profiles (runs $H-J$) as a function of the distance $ r $
  from the macroion center. (a) Fraction $ P(r) $ and (b) radial density $
  n(r) $ of counterions (thin lines) and monomers (thick lines) adsorbed on
  the spherical macroion. (c) Reduced net fluid charge $ Q(r)/Z_{M} $.  }
\label{fig.Pr_nr_Qr_eps80}
\end{figure}

Integrated distribution $ P(r) $, radial distribution $ n(r) $ and fluid net
charge $ Q(r) $ profiles are depicted in Figs. (a-c) respectively. The ion
fraction $ P(r) $ profiles show that for $ f=1 $ (run $H$) and $ f=1/2 $ (run
$I$) almost all atoms lie within a distance $ a<r<20\sigma $ {[}i. e. $
P(r=20\sigma )\approx 1 ${]}, corresponding to roughly one macroion diameter
away from the colloidal surface {[}see Fig. \ref{fig.Pr_nr_Qr_eps80}(a){]}.
This is in contrast to what was previously found (with $ f=1 $ and $ f=1/2 $ )
at stronger Coulomb coupling regimes where almost all ions lie within a
distance of a few monomer sizes from the macroion surface {[}compare Fig.
\ref{fig.Pr_nr_Qr_eps80}(a) with Fig. \ref{fig.Pr_nr_Qr_eps16}(a) and Fig.
\ref{fig.Pr_nr_Qr_eps40}(a){]}. For $ f=1/3 $ (run $J$) only a very small
fraction of monomers {[}$ P_{m}(r=10\sigma )<5\% ${]} lie in the vicinity of
the macroion surface. In this latter situation, the counterion $ P_{c}(r) $
profile indicates that a larger fraction of counterions float in the solution.
Because the $R_g$ of the chain is very large and some chain monomers might be
interacting with the cell boundary we performed for this situation a
simulation where we employed periodic boundary condition, and where the
interactions were computed using the P3M algorithm \cite{deserno98a}. With
this run we found no chain monomers in the vicinity of the colloid surface,
hence unambiguously found monomer desorption from the colloidal surface.

Concerning the intra-chain monomer ordering for $ f=1 $, Fig.
\ref{fig.Pr_nr_Qr_eps80}(b) shows that, although the chain conformation is
relatively dense, there is only one main peak in the monomer $ n_{m}(r) $
profile (the second peak is marginal). This proves that there is no strong
intra-chain monomer ordering (in the normal direction to the macroion sphere)
in this weak Coulomb coupling in contrast with our observations at $
l_{B}=4\sigma $ {[}compare with Fig. \ref{fig.Pr_nr_Qr_eps40}(b){]}.
Nevertheless a second counterion layer (third monomer layer) is build.  The
height $ h_{m} $ of the monomer peak is identical (within the statistical
uncertainty) for $ f=1 $ and $ f=1/2 $ and corresponds to $ h_{m}\approx
0.038/\sigma ^{3} $, whereas for $ f=1/3 $ we have $ h_{m}\approx 0.01/\sigma
^{3} $.

As far as the net fluid charge $ Q(r) $ is concerned, Fig.
\ref{fig.Pr_nr_Qr_eps80}(c) shows that for $ f=1 $ a weak charge oscillation
appears with a marginal macroion overcharge compensation of 3\%. For $ f=1/2
$, the same marginal macroion overcharging occurs but without exhibiting
charge oscillation. Finally, for $ f=1/3 $ no overcharging appears and the the
net charge increases monotonically.

\subsection{Polyelectrolyte chain radius of gyration}

In Fig. \ref{fig.Rg_complex} (for $l_B=2\sigma$) the chain radius of gyration
$ R_{g} $ as function of $ f $ is plotted. It is found that $ R_{g} $
increases almost linearly with $ 1/f $. This result fits well with the scaling
theory in this regime from Ref.\cite{schiessel98a} where it is found that the
chain extension shrinks proportionally with
$l_B$, and we can assume for our purposes $f\propto l_B$. Because the
electrostatic interactions are weaker in the present case ($ l_{B}=2\sigma $),
the ion pair (monomer - condensed counterion) attractions are weaker which
results in a higher $ R_{g} $ value (at fixed $ f $) than in the strong
Coulomb coupling regime.


\subsection{Monovalent case, $f=1$}

Our last result concerns the monovalent case ($ Z_{m}=Z_{c}=1 $
and $ f=1 $) corresponding to run $K$ (see Table \ref{tab.Runs}).
In this situation we have a strong macroion-polyelectrolyte repulsion,
and one can not get like-charge complexation.

\section{Concluding remarks\label{sec.Discussion} }

We have carried out MD simulations to study the complexation of a
charged colloid with a charged polyelectrolyte of the same charge
for various Coulomb couplings $l_B$ and and varying charge fraction $f$. 

For $l_B=10 \sigma$ we gave a reasoning for the observed conformation in terms
of overcharging of the single chain. However this argument only worked for the
largest coupling parameter. 

A complementary view of the observed conformations is to regard both macroions
as being neutralized by their counterions The isolated chain would then
collapse into a globule, and the colloid would be regularly covered by its
counterions.  By changing the Bjerrum length $l_B$ we change the correlations
between the charges which lead in a first approximation to attractions of
dipolar origin, and the attraction is roughly proportional to $l_B$.  On the
other hand, by varying $f$ we change the number of counterions of the chain,
hence the number of available dipoles.

For $f=1$ we can regard the colloidal particle as exerting only a perturbation
of the chain complex (which has a higher density of dipoles). For largest $l_B$
the strong attraction results in a flat disk, for weaker $l_B$ the disk swells
back into the bulk structure, the globule. For smaller $f$ the number of
counterions of the polyelectrolyte and the colloid become comparable in
number, and both macroions can equally compete for the counterions leading to
a greater freedom of both macroions to ``move'' in their common counterion
cloud. For the largest value of $l_B$ we have again the strongest dipolar
attractions leading to a purely 2D conformation, where the chain is wrapped
around the colloid. 
For smaller values of $l_B$ the conformations become more
and more 3D-like, but are still wrapping around the colloid. Of special
interest is here the fact that even for a coupling strength which is typically
for an aqueous solvent ($l_B$ = 7.1 \AA) we find that like-charge complexation
still occurs provided the linear charge density is sufficiently large.
However the adsorption of the polymer chain onto the colloid is weaker than in
the larger Coulomb coupling regimes. For $f=1/2$, we observe formation of
large loops.

For $l_B=2 \sigma$ and $f=1/3$ we find no complex, but end with two single
macroions together with their counterion cloud which interact mainly with
their bare charge, i.e. repulsive.

Our parameters concerning aqueous solutions should be experimentally
accessible, typically for small highly charged colloids (micelles and
relatively short polyelectrolyte chains).

\begin{acknowledgments}
This work is supported by \textit{Laboratoires Europ\'{e}ens Associ\'{e}s}
(LEA).
\end{acknowledgments}


\begin{thebibliography}{10}
\bibitem{NOTE_Many_Body}
This argument assumes that the ions (charged monomers + condensed conterions)
  distribution along the chain is uniform. This should be relatively well
  satisfied since the bare charged monomers are already uniformly distributed.
\bibitem{NOTE_Many_Body_b}
This reasoning holds in the strong Coulomb coupling regime where the poymer
  strucure is essentially dictated by the charged (effective) monomers.
\end{thebibliography}

\end{document}